\documentclass[english]{iopart}
				   
\begin{document}
				   
\title[Holonomy of the Ising model form factors] 
{\Large
 Holonomy of the Ising model form factors
}

\author{S. Boukraa$^\dag$,  S. Hassani$^\S$,
J.-M. Maillard$^\ddag$, B. M. McCoy$^\P$,
W. P. Orrick$^\star$ and N. Zenine$^\S$}
\address{\dag Universit\'e de Blida, LPTHIRM and
 D\'epartement d'A{\'e}ronautique,
 Blida, Algeria}
\address{\S  Centre de Recherche Nucl\'eaire d'Alger, \\
2 Bd. Frantz Fanon, BP 399, 16000 Alger, Algeria}
\address{\ddag\ LPTMC, Universit\'e de Paris 6, Tour 24,
 4\`eme \'etage, case 121, \\
 4 Place Jussieu, 75252 Paris Cedex 05, France} 
\address{\P Institute for Theoretical Physics,
State University of New York,
Stony Brook, USA}
\address{$\star$ Department of Mathematics, 
Indiana University, Bloomington, Indiana 47405, USA} 
\ead{maillard@lptmc.jussieu.fr, maillard@lptl.jussieu.fr, 
mccoy@max2.physics.sunysb.edu,
boukraa@mail.univ-blida.dz, njzenine@yahoo.com, worrick@indiana.edu}

\begin{abstract}
We study the Ising model two-point diagonal correlation function $\, C(N,N)$
by presenting an exponential and form factor expansion 
in an integral representation which differs
from the known expansion of Wu, McCoy, Tracy and Barouch. We extend
 this expansion, weighting, by powers of a variable $\lambda$,
 the $j$-particle contributions, $\, f^{(j)}_{N,N}$. The
 corresponding $\, \lambda$ extension of the two-point diagonal
 correlation function,  $\, C(N,N; \lambda)$, is shown, for 
arbitrary $\lambda$,  
to be a solution of the sigma form of the Painlev{\'e} VI 
equation introduced by Jimbo and Miwa.  Linear 
differential equations for the form factors $\, f^{(j)}_{N,N}$ 
are obtained and shown to have both a ``Russian doll'' nesting, 
and a decomposition of the 
differential operators as a direct sum of operators equivalent
 to symmetric powers of the differential operator
of the elliptic integral $\, E$. Each $\, f^{(j)}_{N,N}$ is expressed 
polynomially in terms of the complete 
elliptic integrals $\, E$ and $\, K$. The scaling limit 
of these differential operators breaks the direct sum 
structure but not the  ``Russian doll'' structure. The 
previous $\, \lambda$-extensions, $\, C(N,N; \, \lambda)$ are, 
for singled-out values  
$\, \lambda=\, \cos(\pi m/n)$ ($m, \, n$ integers), also
 solutions of linear differential equations. 
These solutions of Painlev\'e VI  are actually algebraic functions, being 
associated with modular curves.

\end{abstract}

\vskip .1cm

\vskip .2cm
\noindent {\bf AMS Classification scheme numbers}: 
33E17, 33E05, 33Cxx,  33Dxx, 14Exx, 14Hxx, 34M55, 47E05,
 34Lxx, 34Mxx, 14Kxx
\vskip .1cm
 {\bf Keywords}: form factors, sigma form of Painlev\'e VI, two-point 
correlation functions of the Ising model,  
Fuchsian linear differential equations,  
 complete elliptic integrals, elliptic 
representation of Painlev\'e VI, scaling limit of the Ising
model, algebraic solutions of Painlev\'e VI, 
modular curves, Eisenstein series, quasi-modular forms.

\section{Introduction}
\label{intro}

The two dimensional Ising model in zero magnetic
 field is, historically, the most
important solvable model in all of theoretical physics. The free
energy~\cite{ons1}, the partition function on the finite lattice~\cite{kauf} 
and the spontaneous magnetization~\cite{ons2,yang} were computed 
long ago by Onsager, Kaufman and Yang. These computations, and
subsequent studies of the correlation
 functions~\cite{ko}-\cite{wu-mc-tr-ba-76}, 
 form the basis of 
scaling theory and of the renormalization group approach to critical
phenomena.

The next most important macroscopic property of the Ising model, which
one would like to compute, is the magnetic susceptibility at zero
magnetic field,
which is expressed in terms of the two-point
 correlation functions $\, C(M,N)$ with a
spin at the origin and the other spin in row $\, M$ and column $\, N$, as
\begin{eqnarray}
\label{series}
k_B\,T \cdot \chi \, =  \, \, \sum_{M,N=-\infty}^{\infty} \Bigl(
C(M,N) \, -{\cal M}^2(0) \Bigr)
\end{eqnarray}
where $\, {\cal M}(0)$ is the spontaneous magnetization (which is only non zero
for $\, T<T_c$). Unlike the free energy, and spontaneous magnetization,
this has no known closed form expression, and the study of the magnetic
susceptibility has been the most challenging outstanding problem in the field
for over 50 years.

The first serious analytic study of the susceptibility was made in
1976 by Wu, McCoy, Tracy and Barouch~\cite{wu-mc-tr-ba-76} 
who used their expansions
of the correlation functions to write
the susceptibility as an infinite series in multiparticle
contributions as
\begin{eqnarray}
\chi_{\pm}(T) \,=\,\, \sum_{j}^{\infty}\,\chi^{(j)}
\end{eqnarray}
where the subscript $\pm$ refers to $\,T$ above (resp. below)
 $\,T_c$ and the sum
is over odd (resp. even) values of $j$ for $\,T$ above (below) $\,T_c$. In
ref. \cite{wu-mc-tr-ba-76} the contributions $\,\chi^{(1)}$ 
and $\,\chi^{(2)}$ were explicitly calculated.

No further analysis of  the  susceptibility $\,\chi$,
or of the $\,\chi^{(j)}$, was attempted
 until 1999 when Nickel, in
two remarkable papers~\cite{nickel1,nickel2}, 
showed for $\, j\,\geq \, 3$ that the $\,\chi^{(j)}$'s
have singularities in the complex temperature plane whose
number increases with $\, j$ and become dense on a circle 
as $\, j\, \rightarrow \, \infty$. 
Unless a remarkable cancellation takes place this discovery implies
that the magnetic susceptibility will have a natural boundary in the
complex temperature plane which extends to $\, T_c$. This natural
boundary is a new phenomenon which is
not incorporated into scaling, or renormalization theory, and, thus, it raises
significant questions about our understanding of critical phenomena.
Consequently it is most important to deeply understand the properties of the
Ising susceptibility, and this challenging question certainely requires
 some serious progress on the two-point correlation functions of the Ising
model.  
Note that some re-summed high temperature series \cite{hanseletal, Enting}
in the anisotropic case has, already, enabled Guttmann and Enting~\cite{Enting}
 to conjecture, for the anistropic $\chi$, a natural boundary 
in one variable when the second variable is fixed.

In 2001 the work of Orrick {\it et al.}~\cite{or-ni-gu-pe-01b} provided 
a polynomial time algorithm for 
obtaining the coefficients of the susceptibility series
 of the two-dimensional Ising model:
from a combinatorial enumerative viewpoint this
 can be viewed as a ``solution'' of the 
problem. The existence of such a polynomial time algorithm for a lattice 
problem, instead of the exponential growth of the
 calculations one expects at first sight,
 can be seen as some ``combinatorial integrability'' 
of the model~\cite{Enting}. 
However a (very) efficient way of getting very large series expansions 
for a physical quantity of a model of lattice statistical
 mechanics, is far from
 providing the closed formula and exact results
 one might desire : for instance, there is still a lot
 of work to be done in order to
extract singular points, singular behaviours, 
from the knowledge of very large series.  

In 2004 several of the present
authors~\cite{ze-bo-ha-ma-04,ze-bo-ha-ma-05,ze-bo-ha-ma-05b,ze-bo-ha-ma-05c}
 initiated the study of the
Ising susceptibility, beyond the singularity analysis 
of Nickel~\cite{nickel1,nickel2},
 by determining the Fuchsian linear differential
equations for $\, \chi^{(3)}$ and $\,\chi^{(4)}$ as a function of the
temperature. These equations have many
remarkable properties such as a ``Russian doll'' nesting structure :
 the function $\,\chi^{(1)}$ satisfies the equation for $\chi^{(3)}$
and $\,\chi^{(2)}$ satisfies the equation for $\,\chi^{(4)}$. If this
nesting can be proven to extend to all of the $\chi^{(j)}$'s  there must
be remarkable structures in the Fuchsian equations and the hope is thus
raised that it may be possible to 
characterize\footnote[3]{The full susceptibility could be the solution of
a nonlinear equation, or the solution of a system of PDE's, or 
solution of a nonlinear
 functional equation, or ...} the full susceptibility.

In a more recent paper~\cite{PainleveFuchs} several of the
 present authors provided 
new results on the exact expressions of the 
two-point correlation functions of the Ising model,
especially the  diagonal correlation $\,C(N,N)$,
 underlining the key role played by the second order linear
 differential operator
corresponding to the complete elliptic integral
 of the first or second kind $\,K$ or $\,E$.

In this paper, we study the diagonal correlation functions
$\,C(N,N)$ as a form factors expansion.
Our starting point will be the expansions of the diagonal correlations
in an exponential form~\cite{wu-mc-tr-ba-76}, both for $\,T<T_c$
\begin{eqnarray}
C_{-}(N,N)\, =\, \, \, 
(1-t)^{1/4}\cdot  {\rm exp} \Bigl(\sum_{n=1}^{\infty}\, F^{(2n)}_{N,N} \Bigr)
\label{cm}
\end{eqnarray}
with
\begin{eqnarray}
t\,=\,\, \Bigl( \sinh(2E^v/k_BT)\sinh(2E^h/k_BT) \Bigr)^{-2}
\end{eqnarray}
and for $T>T_c$
\begin{eqnarray}
\label{cp}
C_{+}(N,N)\, =\,\,\,  
 (1-t)^{1/4} \cdot  \sum_{n=0}^{\infty}\,
 G^{(2n+1)}_{N,N} \cdot {\rm exp}\,
 \Bigl( \sum_{n=1}^{\infty}\, F^{(2n)}_{N+1,N+1} \Bigr)
\end{eqnarray}
with
\begin{eqnarray}
t\, =\,\, \Bigl( (\sinh(2E^v/k_BT)\sinh(2E^h/k_BT) \Bigr) ^2
\end{eqnarray}
where $\, E^h$ and  $\, E^v$ are the horizontal and vertical interaction 
energies of the Ising model.
When the exponentials in (\ref{cm}) and (\ref{cp}) are expanded,
the correlations 
 can also be written in what is called  a ``form factor'' expansion :
\begin{eqnarray}
\label{formm}
&&C_{-}(N,N)\, =\, \, \,(1-t)^{1/4} \cdot
\Bigl(1+\sum_{n=1}^{\infty}\, f^{(2n)}_{N,N}\Bigr)  \\
&&C_{+}(N,N)\, =\, \,  
(1-t)^{1/4} \cdot \sum_{n=0}^{\infty}\, f^{(2n+1)}_{N,N} 
\label{formp}
\end{eqnarray}
The form factor $\,f^{(j)}_{N,N}$ is interpreted as the ``$j$-particle'' 
contribution  to the two-point correlation function.
 It is natural to consider
$\, \lambda$-extensions~\cite{wu-mc-tr-ba-76,mtw} of the previous functions
\begin{eqnarray}
\label{formm1}
&&C_{-}(N,N;\lambda)\, = \, \,\, (1-t)^{1/4} \cdot
\Bigl(1+\sum_{n=1}^{\infty}\lambda^{2n} f^{(2n)}_{N,N}\Bigr) \\
&&C_{+}(N,N;\lambda)\, =\, \,\,\, (1-t)^{1/4} \cdot 
\sum_{n=0}^{\infty} \, \lambda^{2n} \cdot  f^{(2n+1)}_{N,N}  \label{formp1}
\end{eqnarray}
which weight each $\,f^{(j)}_{N,N}$ by some power
 of $\,\lambda$, and to interpret $\,\lambda$
as being analogous to a coupling constant in a quantum field theory expansion.
Such $\, \lambda$-extensions naturally 
emerge from the {\em Fredholm determinant
framework} in~\cite{wu-mc-tr-ba-76}.
We will present new integral representations 
 for $\, F^{(2n)}_{N,N},~G^{(2n+1)}_{N,N}$ and
 $\, f^{(j)}_{N,N}$ in sec.(\ref{new}). We will see that they are much simpler,
  and more transparent, than the forms obtained
 from $\, C(M,N)$ of~\cite{wu-mc-tr-ba-76}
  by specializing to $\, M=\, N$. The proof of these results is obtained by
  extending the expansion solution for the leading term given in 1966
  by Wu~\cite{wu}, to all orders. It  will be published elsewhere.

The diagonal correlations $\,C(N,N)$ have the property, discovered by
 Jimbo and Miwa~\cite{jm} in 1980, that their log-derivatives are solutions of 
the ``sigma'' form\footnote[2]{We use 
a variable $\, t$ which is the inverse  of the one of 
Jimbo and Miwa~\cite{jm}.}
 of a Painlev{\'e} VI function
\begin{eqnarray}
\label{jimbo-miwa}
&&\left( t\, (t-1)\,  {d^2\sigma \over dt^2} \right)^2 \, 
\, = \, \,  N^2 \left( (t-1) {d\sigma \over dt} -\sigma \right)^2\,
\nonumber \\
&& \qquad  \quad -4 {d\sigma \over dt} \left((t-1) {d\sigma\over dt}
-\sigma \,  -{1\over  4} \right) \left(t {d\sigma \over dt}\,  -\sigma \right) 
\end{eqnarray}
where $\, \sigma \,$ is defined for $\, T\,<\, T_c \, $ as
\begin{eqnarray}
\label{sigmam}
\sigma_N(t)\,\, =\,\,\,
 t\, (t-1)\cdot {d \ln C_{-}(N,N)\over dt}\,\, -{t\over 4}
\end{eqnarray}
with the normalization condition
\begin{eqnarray}
\label{cmnorm}
C_{-}(N,N)\,=\,1\, +O(t) \quad {\rm for} \quad \quad t\rightarrow 0 
\end{eqnarray}
and, for $\, T > T_c$, as
\begin{eqnarray}
\label{sigmahigh}
\sigma_N(t)\, =\,\, \,\, 
t\, (t-1)\cdot {d \ln C_{+}(N,N) \over dt}\,\,\,  -{1\over 4}
\end{eqnarray}
with the normalization condition
\begin{eqnarray}
\label{cpnorm}
C_{+}(N,N)\, \,= \, \,\, 
{(1/2)_N\over N!} \cdot t^{N/2}\cdot \left(1+O(t)\right) 
 \quad \quad {\rm for} \quad t\rightarrow 0
\end{eqnarray}
where $\,\,(a)_N\,=\, \Gamma(a+N)/\Gamma(a)\, $ denotes the Pochhammer symbol.

One can easily verify that (\ref{jimbo-miwa}), the $\, N$-dependent
 sigma form of Painlev\'e VI, is actually 
covariant by the  Kramers-Wannier duality :
\begin{eqnarray}
\label{Kramers}
(t, \, \sigma, \,  \sigma', \,  \sigma'') \quad \rightarrow \quad \quad 
\Bigl( {{1} \over {t}},  \, \, \, \, {{\sigma} \over {t}},
 \, \,\, \,   \sigma \, - t \cdot  \sigma', 
\,  \,\, \,  t^3 \cdot  \sigma''
\Bigr)
\end{eqnarray}

On another hand, Jimbo and Miwa introduced in~\cite{jm} 
an {\em isomonodromic} $\, \lambda$-extension of $\, C(N,N)$
 and showed that this more general function $\, C(N,N; \, \lambda)$
also satisfies (\ref{jimbo-miwa}). The motivation of 
introducing an isomonodromic parameter $\, \lambda$,
in the framework of {\em isomonodromy deformations},
 is, at first sight, quite different from the ``coupling constant'' 
 motivation at the origin
of the form factor $\, \lambda$-extensions (\ref{formm1}) and (\ref{formp1}). 
In sec.(\ref{series1}) we show that these two $\, \lambda$-extensions
 are {\em actually the same} by demonstrating that   
the recursive solutions of (\ref{jimbo-miwa}),
analytic\footnote[5]{
The $\, \lambda$-extensions (\ref{formm1}) and (\ref{formp1}) are
analytic at $\, t\, \sim \,0$
in $\, t$ for $\, T \, < \, T_c$ and, when
 $\, T \, > \, T_c$, analytic in $\, t$ for $\, N$ even, and
in  $\, t^{1/2}$ for $\, N$ odd.} in  $\, t^{1/2}$,
agree with (\ref{formm1}) and 
(\ref{formp1}) where the $\, f^{(j)}_{N,N}$'s are
 obtained from $\, C_{\pm}(N,N; \lambda)$,
 the $\, \lambda$ expansion of $\, C_{\pm}(N,N)$ of sec.(\ref{new}).  
The normalization condition (\ref{cmnorm})  fixes
 one integration constant in the solution to (\ref{jimbo-miwa}). 
 We find that the second integration constant is
 a free parameter, and, denoting that parameter
 by $\, \lambda$,  that our one parameter family 
of solutions for $\, C_{-}(N,N)$ can be written
 in a form structurally similar to the 
right hand side of (\ref{formm1}).  Furthermore, we have confirmed,
 by comparison with series expansions of the multiple 
integral formulas for $\, f_{N,N}^{(j)}$
 derived in sec.(\ref{new}), that this family of
 solutions is, in fact, identical to $\, C_{-}(N,N;\lambda)$
 as defined in (\ref{formm1}). Similarly, the
 condition (\ref{cpnorm}) gives rise to a one parameter
 family of solutions for $\, C_{+}(N,N) $ that is 
identical to (\ref{formp1}).  After all, the fact 
that these two distinct $\, \lambda$-extensions of $\, C_{\pm}(N,N) $ identify
 is not altogether surprising, since Jimbo and Miwa's derivation of
 (\ref{jimbo-miwa}) also starts from a multiple-particle expansion
 of the correlation functions in terms of free fermion operators. 
 It does not, however, appear to have been observed previously.

In sec.(\ref{holo}) we use formal computer algebra  to 
study the functions $\, f^{(j)}_{N,N}$. We
 obtain the Fuchsian linear differential 
equations satisfied by the $\, f^{(j)}_{N,N}$ 
 for fixed $\, j \le \, 9$ and arbitrary $\, N$.
We also find the truly remarkable result that the families
 $\, f^{(2j+1)}_{N,N}$ and $\, f^{(2j)}_{N,N}$ are each 
annihilated by linear differential operators which have
 a nested ``Russian doll'' structure. Beyond this  ``Russian doll'' 
structure, each linear differential 
operator  is the {\em direct sum}  of linear differential  operators
equivalent\footnote[8]{For the 
equivalence of linear differential 
operators, see~\cite{Singer,hoeij2,PutSinger}.}
to symmetric powers of the
 second order differential operator corresponding
 to $\, f^{(1)}_{N,N}$, (or equivalently to the
 second order differential operator $\, L_E$,
 corresponding to the complete elliptic 
integral $\, E$).
A direct consequence is that the form factors 
$\, f^{(2j+1)}_{N,N}$, and $\, f^{(2j)}_{N,N}$
are {\em polynomials in  the complete elliptic integrals of
the first and second kinds,} $\, K$ and $\, E$:
\begin{eqnarray}
\label{EK}
K\,=\,\, {_2}F_1 \left( 1/2, 1/2; 1; t \right), \quad \quad
E\,=\,\, {_2}F_1 \left( 1/2, -1/2; 1; t \right)
\end{eqnarray}
A simple example is 
$\, f^{(2)}_{0,0} \,= \, \,  K \cdot (K-E)/2$.

The closed formula we obtain for the differential operators in these
nested ``Russian doll'' structures, enable us to take the 
{\em scaling limit} of these operators.
We study this scaling limit in sec.(\ref{scal})
and show that the  ``Russian doll'' structure remains valid. 
The differential operators
in that ``scaled'' nested Russian doll structure remain equivalent to
 the symmetric power of a singled out second order
differential operator (corresponding to the modified Bessel function).
In contrast, in the scaling
 limit, {\em the direct sum of operators decomposition 
structure is lost}, 
and we explain why.

The unexpectedly simple expressions
for the form factors $\, f^{(j)}_{N,N}$ of
 sections (\ref{new})--(\ref{scal}), and the corresponding
remarkable  differential structures, 
may be used to obtain many further results.
We display some of these results in sec.(\ref{algebr}).
Recalling that, when $\lambda=\, 1$, the Ising 
correlation functions $\, C(N,N;\, 1)$ satisfy
Fuchsian differential equations~\cite{PainleveFuchs} with an order that 
grows with $\, N$,  it is quite natural to inquire
 whether there are any other values of $\, \lambda$
for which $\, C(N,N;\lambda)$ will satisfy a 
Fuchsian linear differential  equation. One such
family of $\, \lambda$ is motivated
 by the work of Cecotti and  Vafa~\cite{cv} on
$\, N\, =2$ supersymmetric field theories where they encountered 
$\lambda$ extensions of the Ising correlations in the scaling limit~\cite{mtw}
with ($m\, $ and $\, n$ are integers)
\begin{eqnarray}
\label{root}
\lambda \, = \, \, \, \cos(\pi m/n)
\end{eqnarray}
Indeed, we have found that for
$ \, n \, = \,\,  3,\cdots, \, 20$,  the functions $ \, C(N,N;\lambda)$
satisfy  Fuchsian linear differential equations
 whose orders, in contrast with those of the
$\, \lambda=\, 1$ equations~\cite{PainleveFuchs},
 {\em do not} depend on $\, N$.  More
importantly, we find that these solutions
 are {\em actually algebraic functions} of
 $\, t$, associated with {\em modular curves}. 

We conclude, in sec.(\ref{concl}), with a discussion about the significance
of our results on the factorization of multiple dimensional integrals. 

\section{New integral representations for  the $\, f^{(n)}_{N,N}$'s }
\label{new}

The form factor expressions for $\, C(M,N)$ 
of~\cite{wu-mc-tr-ba-76,nickel1, nickel2, pt,yamada,or-ni-gu-pe-01b} 
are obtained by expanding the exponentials in (\ref{cm}), and (\ref{cp}), 
in the form given in \cite{wu-mc-tr-ba-76}  as multiple integrals 
and integrating over half the variables. The form of the result depends 
on whether the even, or odd, variables of~\cite{wu-mc-tr-ba-76} are integrated 
out. For the general anisotropic lattice, one form of this result   is
given,  for arbitrary $\, M$ and $\, N$, in~\cite{or-ni-gu-pe-01b}.
When specialized to the isotropic case the result is
\begin{eqnarray}
\label{fnormali}
 f^{(2j)}_{M,N}  \, = \, \,  \hat{C}^{2j}(M, \, N), \qquad
 f^{(2j+1)}_{M,N}  \, = \, \,  {{\hat{C}^{2j+1}(M, \, N)} \over {s}} \qquad
\end{eqnarray}
where $\, s$ denotes $\, \sinh(2\, K)$, and where 
\begin{eqnarray}
\label{involved}
&&\hat{C}^{j}(M, \, N) \, = \, \, \,
 \, {1 \over {j!}} \, 
\int_{-\pi}^{\pi}\, {{d\phi_1} \over {2 \, \pi}} \, \cdots \, 
\int_{-\pi}^{\pi}\, {{d\phi_j} \over {2 \, \pi}} \, 
\Bigl(  \prod_{n=1}^{j} \, {{1} \over {\sinh \gamma_n}} \Bigr) \nonumber \\
&& \qquad \times  \, 
\Bigl(  \prod_{1 \le i \le k \le j} \, h_{ik} \Bigr)^2 \,
 \Bigl( \prod_{n=1}^{j} \, x_n \Bigr)^M \, 
\cos\Bigl( N \, \sum_{n=1}^{j} \,\phi_n \Bigr)
\end{eqnarray}
with :
\begin{eqnarray}
&&x_n \,\,\, = \, \,\, \,
s \, + {{1} \over {s}} \, -\cos \phi_n \,
 -\Bigl( ( s \, + {{1} \over {s}}  -\cos \phi_n )^2 \, -1 \Bigr)^{1/2}, \\
&& \sinh \gamma_n \, \,= \,\, \,
\Bigl(  ( s \, + {{1} \over {s}}  -\cos \phi_n )^2 \, -1 \Bigr)^{1/2},  \\
&&  h_{ik}\,\, = \,\,\, {{ 2 \, (x_i\, x_k)^{1/2} \, 
 \sin((\phi_i - \phi_k)/2)  } \over {1 \, -x_i\, x_k }}
\end{eqnarray}

In this work, we obtain the expressions of  $\, f^{(j)}_{N,N}$
not by setting $\, M=\, N$ 
in the results of~\cite{wu-mc-tr-ba-76}, but, rather, 
from the representations of $\, C(M,N)$ as an $\, N$-dimensional
 Toeplitz determinant with elements
\begin{eqnarray}
\label{ammoinsn}
&&a_{m,n}\,  = \,\,a_{m-n}\,= \\
&&\qquad \quad \,\,{\frac{{1}}{{2\,\pi }}}\cdot \int_{0}^{2\, \pi}
\,d\theta \,e^{-i\,(m-n)\theta }\,\Bigl({\frac{{(1\,-\alpha
_{1}\,e^{i\,\theta })(1\,-\alpha _{2}\,e^{-i\,\theta })}}{{(1\,-\alpha
_{1}\,e^{-i\,\theta })(1\,-\alpha _{2}\,e^{i\,\theta })}}}\Bigr)^{1/2}
 \nonumber
\end{eqnarray}
with
\begin{eqnarray}
\label{replaceCNN}
\alpha _{1}\,=\,0 
 \quad \quad \hbox{ and } \quad  \quad \alpha _{2} \, = \, s^{-2}
\end{eqnarray}
for the diagonal correlation $C(N,N)$, and
\begin{eqnarray}
\label{replaceC0N}
&&\alpha _{1}\,=\, \Bigl( (1+s^2)^{1/2} \, -s \Bigr)\cdot \Bigl( 
 {{(1+s^2)^{1/2} \, -1 } \over {s}} \Bigr) 
 \quad  \quad \hbox{ and } \nonumber \\
&& \quad \alpha _{2}\, =\,
\, \Bigl( (1+s^2)^{1/2} \, -s\Bigr) \cdot 
\Bigl(  {{(1+s^2)^{1/2} \, +1 } \over {s}} \Bigr) 
\end{eqnarray}
for the row correlation\footnote[2]{Although in this paper 
we take the particular values of $\, \alpha _{1} $ and $\, \alpha _{2} $
corresponding to $\, C(N,N)$, our results are also, {\em mutatis mutandis,} 
applicable to the correlations  $\, C(0,N)$ and to the triangular lattice with 
$\, \alpha _{1} $ and $\, \alpha _{2} $ given by~\cite{Stephenson}.
} $\, C(0,\, N)$.
Our method  is to follow Wu's paper~\cite{wu}
in the framework of the general theory of Toeplitz
 determinants.

For $ \, T < \, T_c$, let us first recall (3.15) 
of Wu's paper~\cite{wu}, which reduces, for the 
 diagonal correlations $\, C(N,N)$,
to\footnote[5]{To be precise,
note that Wu considered in his paper, the  $\, C(0,N)$ correlations.
From the definition (\ref{ammoinsn}) of the entries in the Toeplitz determinant
one can consider the diagonal correlations $\, C(N,N)$ with the replacement
 (\ref{replaceCNN}) instead of (\ref{replaceC0N}).}:
\begin{eqnarray}
\label{315}
&&(1-t)^{-1/4} \cdot  C(N,N)
 \, \, \,\sim \, \,\, \, \nonumber \\
&&\quad \quad \quad 1+ \, {{1} \over {(2\, \pi)^2}} 
\, \int d\xi \cdot  \xi^N \cdot 
\Bigl((1-\alpha_2\, \xi)  (1-\alpha_2/\xi) \Bigr)^{-1/2} \,    \\
&& \quad\quad \quad \quad \times \, \int d\xi' \cdot \xi'^{-N} \cdot
\Bigl((1-\alpha_2\, \xi')  (1-\alpha_2/\xi') \Bigr)^{1/2} 
 {{ 1} \over {(\xi'\, -\, \xi)^2  }} \nonumber 
\end{eqnarray}
Comparing with (\ref{formm}) we see that the second term in
(\ref{315}) is $\,\,f^{(2)}_{N,N}\,=\,\, F^{(2)}_{N,N}$.

Performing the change of variables $\, \xi\, = \, z_1$ 
and $\, \xi'\, = \, 1/z_2$,
deforming the contour of integration for both $\, z_1$ and $\, z_2$
(one has to consider only the discontinuity 
across the branch cut\footnote[3]{For $\, T<T_c$,
 $\, \alpha_2\, = \, t^{1/2}\, < \, 1$.  } 
running from $\, 0$ to $\, \alpha_2$),
and rescaling $\, z_1$ and $\, z_2$, in, respectively,
 $\, x_1\, = \,z_1/\alpha_2 $
and $\, x_2\, = \, z_2/\alpha_2$, we obtain :
\begin{eqnarray}
\label{f2}
&&f^{(2)}_{N,N}(t) \, \, = 
\,\, \,F^{(2)}_{N,N}(t) \, \,= \,\, \,\,
      {t^{(N+1)}\over
  \pi^2}\, \int_0^1\,  x_1^N\,  dx_1 \, \int_0^1\,  x_2^N\,  dx_2 \nonumber \\
&& \qquad \qquad \quad  \times \left({x_{1}(1-x_{2})(1\, -t\, x_{2})\over 
 x_{2}(1-x_{1})(1\,-t\, x_{1})}\right)^{1/2}  \,
  (1\,-t\,x_{1}\,x_{2})^{-2}\,
\end{eqnarray}

Similarly, when $\, T >T_c$, the leading term for  $\, G_{N, \, N}^{(1)}  $
is given by equation (2.29) of~\cite{wu}:
\begin{eqnarray}
\label{G0N1}
 f_{N, \, N}^{(1)}  \, = \, \, G_{N, \, N}^{(1)}  \,= \, \,
{{-1} \over {2 \, \pi \, i}}
 \, \int_{C} \, dz \, {{ z^{N-1} } \over { 
\Bigl( (1\, - t^{1/2}\, z )(1\, -\, t^{1/2}\, z^{-1}) \Bigr)^{1/2} }}
 \quad  \quad 
\end{eqnarray}
which, after deforming the contour of integration to the branch cut,
and scaling   $\, z \, = \,  t^{1/2} \, x $, becomes
\begin{eqnarray}
\label{1}
&&f_{N,N}^{(1)}(t)\,\,=\,\,\,G_{N,N}^{(1)}(t)\,\,\nonumber  \\
&& \quad \quad\quad =\,\,\,{{t^{N/2}} \over {\pi}} \cdot 
 \int_0^{1}  \, x^{N-1/2} \, (1-x)^{-1/2} \, (1\, -x\, t)^{-1/2} \, dx 
\nonumber\\
 &&\quad \quad \quad =\,\,
 t^{N/2} \cdot {(1/2)_N\over N!}  \cdot
 {_2}F_1\Bigl({1\over 2},N+{1\over 2};\, N+1;t \Bigr)
\end{eqnarray}
where $\,\,\, {_2}F_1(a,b;c;z)\,\, $ is the
 hypergeometric function~\cite{Erde}.

 The full expressions for  $\, F_{N, \, N}^{(2n)}$ for $ T < T_c$,  and 
 $\, F_{N+1, \, N+1}^{(2n)}$
and $\,   G_{N, \, N}^{(2n+1)} $ for $ T > T_c$, 
can be obtained by following the iterative procedure 
based on (2.9)--(2.16) of~\cite{wu-mc-tr-ba-76} 
to all orders, just as the full expressions for 
$\, F_{ M, \, N}^{(2n)}$
and $\,   G_{M, \, N}^{(2n+1)} $ with $\, M \, \ne 0$
of (2.9)-(2.16) of~\cite{wu-mc-tr-ba-76} 
are obtained, in sections 3 and 4 of~\cite{wu-mc-tr-ba-76},
by following the procedure of Cheng and Wu~\cite{cw}
to all orders\footnote[4]{The full expressions for  $\, F_{ 0, \, N}^{(2n)}$
and $\,   G_{0, \, N}^{(2n-1)} $
can also be obtained by performing the procedure based on (2.9)--(2.16)
of~\cite{wu-mc-tr-ba-76} to all orders just as the full expressions for 
$\, F_{ M, \, N}^{(2n)}$
and $\,   G_{M, \, N}^{(2n-1)} $ with $\, M \, \ne 0$
of (2.9)-(2.16) of~\cite{wu-mc-tr-ba-76} are obtained in
 sections 3 and 4 of~\cite{wu-mc-tr-ba-76},
by ``cycling'' the procedure of Cheng and Wu~\cite{cw}
to all orders.}. The details of the ``iterative procedure''
 will be presented elsewhere\footnote[2]{The first step
 in that calculation is to
consider the ratio $\, C_{+}(N,N, t)/C_{-}(N,N, t)$. This
 will be detailed elsewhere.}.
 These are certainly implicit in the paper of Jimbo and Miwa~\cite{jm},
however, we have not been able to find a reference where they are explicitly
written out.

When the low temperature 
expansion of sec. 3 of Wu~\cite{wu} is performed  
to all orders, we find that (94) holds with
\begin{eqnarray}
&&F^{(2n)}_{N,N} \, =\, \, \,\,  
{(-1)^{n+1} \over n}{1 \over (2\pi)^n}\int \prod_{j=1}^{2n}
{z_j^N\,  dz_j \,   \over 1\,-z_jz_{j+1}} \\
&&\quad \quad \quad \quad \quad \quad \times \prod_{j=1}^{n}\left({
  (1-\alpha_2\, z_{2j})(1-\alpha_2/z_{2j}) \over
(1 \, -\alpha_2 \, z_{2j-1})(1 \, -\alpha_2/z_{2j-1})}\right)^{1/2}
\nonumber 
\end{eqnarray}
from which, after deformation of integration contours and rescaling,
one obtains, for $\, T < \, T_c$,
the following  {\em new integral representation} of $ \, F^{(2n)}_{N,N}(t)$ :
\begin{eqnarray}
\label{Fn}
&& F_{N,N}^{(2n)}(t)\,\,\, \,  =\, \, 
  {(-1)^{n+1}\, t^{n(N+1)}\over n \, \pi^{2n}} \times  \\
&&\quad \times \int_{0}^1
  \prod_{j=1}^{2n}{\,x_j^N \, dx_j \over 1\,-t\,x_j\,x_{j+1}} \cdot 
\prod_{j=1}^{n}\, \left({x_{2j-1}(1\, -x_{2j})(1-tx_{2j})
\over x_{2j}(1\, -x_{2j-1})(1\, -t\,x_{2j-1})}\right)^{1/2}\nonumber 
\end{eqnarray}

Similarly for $\,T\,>\,T_c\,$ the expansion of sec.2 of Wu~\cite{wu} 
is performed to 
all orders and we find that (\ref{cp}) holds with $\,F^{(2n)}_{N,N}\,$
 given by (\ref{Fn}) and
\begin{eqnarray}
&&G^{(2n+1)}_{N,N}\, =\, \, \, \, 
 (-1)^n \, {1\over (2\pi)^{2n+1}}\nonumber  \\
&&\quad \quad \times \int  \prod_{j=1}^{n+1}(z_{j}^{N+1} dz_{j} )\,
 {{ 1} \over {z_1\, z_{2n+1} }}
\,  \prod_{j=1}^{2n}{1\over 1\, -z_jz_{j+1}}\nonumber\\
&&\quad \quad  \times \prod_{j=1}^{n+1}\left(
(1\, -\alpha_2^{-1}z_{2j-1})
(1\, -\alpha_2^{-1}/z_{2j-1})\right)^{-1/2}\nonumber\\
&&\quad \quad \times \prod_{j=1}^{n}
\Bigl((1\, -\alpha_2^{-1}z_{2j})
(1\, -\alpha_2^{-1}/z_{2j})\Bigr)^{1/2} 
\end{eqnarray}
Changing variables and deforming contours, we obtain :
\begin{eqnarray}
\label{gn}
&&G^{(2n+1)}_{N,N}(t)\, = \, \, \, (-1)^{n} \, \,  {t^{N(2n+1)/2+2n} \over
 \pi^{2n+1}} \nonumber \\
&&\quad \quad  \times
 \int_{0}^{1} \, \prod_{j=1}^{2n+1}(x_{j}^{N+1}\, dx_{j})
 \, \, \,{{1} \over {x_1 \,x_{2n+1}}}  \,  \,
\prod_{j=1}^{2\, n}{1\over  1\, -t\, x_j\, x_{j+1}}\nonumber \\
&&\quad \quad \times \prod_{j=1}^{n+1}\left({x_{2j-1}\over
  (1\, -x_{2j-1})(1\, -t\,x_{2j-1})}\right)^{1/2} \nonumber \\
&&\quad\quad  \times \, \, \prod_{j=1}^{n} \Bigl(
  (1\, -x_{2j})(1\, -t\,x_{2j})/x_{2j}\Bigr)^{1/2} 
\end{eqnarray}
 
The form factor expressions are then obtained by expanding the exponentials.
Thus we find, for $\, T<T_c$, that the form factors in (\ref{formm1}) read
\begin{eqnarray}
\label{2n}
&&f_{N,N}^{(2n)}(t)\,=\,  {{  t^{n(N+n)}} \over {
  (n!)^2 }} \, {{1 } \over {\pi^{2n} }} \cdot 
\int_0^1\prod_{k=1}^{2n} \,x_k^N \, dx_k \,\nonumber \\
&&\quad \quad  \times \prod_{j=1}^n \,
 \left({x_{2j-1}(1-x_{2j})(1\, -t\, x_{2j})\over 
x_{2j}(1-x_{2j-1})(1\, -t \,x_{2j-1})}\right)^{1/2} \nonumber\\
&& \quad \quad \times
 \, \prod_{1\leq j\leq n} \,
\prod_{1 \leq k \leq n}(1\, -t\, x_{2j-1}\, x_{2k})^{-2}
\nonumber\\
&&\quad\quad  \times  
\prod_{1\leq j<k\leq n}(x_{2j-1} -x_{2k-1})^2(x_{2j} -x_{2k})^2 
\end{eqnarray}
and, for $\, T>T_c $, the odd form factors in (\ref{formp1}) read 
\begin{eqnarray}
\label{2n-1}
&&f_{N,N}^{(2n+1)}(t)\,=  \,  \,\,\, t^{((2n+1)N/2  \,+n(n+1))}
 \cdot {{ 1 } \over {\pi^{2n+1} }} 
\cdot {{ 1 } \over {n! \, (n+1)! }} \nonumber\\
&&\quad \quad  \times \int_0^1\prod_{k=1}^{2n+1}\, x_k^{N}\, dx_k \, 
\prod_{j=1}^{n+1}\, \Bigl((1-x_{2j})
(1\,-t\,x_{2j})\, x_{2j}\Bigr)^{1/2}\nonumber\\
&&\quad \quad \times
\prod_{j=1}^{n+1}\, 
\Bigl((1-x_{2j-1})(1\,-t\, x_{2j-1})\, x_{2j-1}\Bigr)^{-1/2}
\nonumber\\
&&\quad \quad  \times \prod_{1\leq j\leq n+1}\prod_{1\leq k \leq n}
(1\, -t\, x_{2j-1}\, x_{2k})^{-2}   \,   \\
&&\quad\quad  \times \prod_{1\leq j<k\leq n+1}(x_{2j-1}-x_{2k-1})^2
\prod_{1\leq j<k\leq n}(x_{2j}-x_{2k})^2 \nonumber
\end{eqnarray}
where the last product in (\ref{2n-1}) has to 
be taken to be equal to unity for $ n=\, 0,1$.
We note that the factors $\, 1/(n!)^2\, $ and $\, 1/(n! \, (n+1)!)$
in (\ref{2n}), and  (\ref{2n-1}), arise because the integrands are symmetric 
functions of the variables $\, x_{2j}$ and $\, x_{2j-1}$, separately. This 
is to be contrasted with (\ref{involved}) where there is no separation in the
odd and even integrals $\, \phi_{j}$.

In the simplest case the previous integral representation (\ref{2n-1})
gives  $ \, f_{N,N}^{(1)}(t)$ defined by (\ref{1})
where one recognizes the Euler representation of an hypergeometric function.

Do note that the $\, (G^{(2n+1)}_{N,N}, \, F^{(2n)}_{N+1,N+1})$ decomposition
in (\ref{cp}) {\em is not unique}. In contrast, the form factor expressions
(\ref{2n}), (\ref{2n-1}) are unique and well-defined.

\vskip .1cm

It is tempting to try to ``bridge'' 
such new integral representations
(\ref{2n}), (\ref{2n-1}) 
with integral formulas like (\ref{involved}), or other integral formulas
one can find 
in~\cite{wu,cw,wu-mc-tr-ba-76,nickel1,nickel2,mcc-wu-80}, getting 
 (\ref{2n}), (\ref{2n-1}) from these other integral 
formulas after some changes of variables,
 or from partial integrations on a subset of variables in order to 
reduce  $\, (4n)$ integral formulas 
 into  $\, (2n)$ integrals. We have not been able to do this. 
Basically we have two kinds 
of drastically different formulas: the ones emerging 
from {\em Fredholm determinant} expansions
that naturally yield integral formulas with integrands 
that are algebraic functions of the self-dual
variable $\, w \, = \, s/(1+s^2)/2$, and the ones 
emerging from ``isomonodromic'' calculations\footnote[2]{
For instance formulas similar to  formulas (13) and (14) of~\cite{jm}.}
that  naturally yield integral formulas with integrands 
that are algebraic functions of 
the modulus $\, k$ of the elliptic 
function (or the variables $\, s$ or $\, t$; for $\, T \, > \, T_c$,
 $k\, = \, s^2$,  $  \,t \, = \, s^4$)
 and break the duality $\, s \, \rightarrow \, 1/s$.  
Do note that, in the case of the isotropic
lattice, the formulas in~\cite{wu-mc-tr-ba-76,mcc-wu-80} 
are naturally integral formulas
 with integrands that are algebraic functions of the self-dual
variable $\, w \, = \, s/(1+s^2)/2$. 
{\em It is only in the scaling limit} that these integral formulas
look like  integral formulas with integrands that
 are algebraic functions of $\, s$, or $\, k$, 
or $\, t$. The ``$\, t$-integral formulas''
 of the second kind (\ref{2n}), (\ref{2n-1}) 
naturally produce series expansions in the 
hypergeometric functions $ _{2}F_{1}$, 
while the ``$\, w$-integral formulas'' of the first kind naturally
 generate series expansions in the hypergeometric
 functions $ _{4}F_{3}$ : see for instance, all the
series calculations we
 obtained 
in~\cite{ze-bo-ha-ma-04,ze-bo-ha-ma-05,ze-bo-ha-ma-05b,ze-bo-ha-ma-05c} 
in the holonomic analysis of $\, \chi^{(3)}$ and $\, \chi^{(4)}$.
The  $ _{4}F_{3}$ we consider are particular
and, consequently, can be written, for fixed parameters, 
in terms of $ _{2}F_{1}$ for which 
quadratic transformations\footnote[3]{For $\, a=b=1/2\, $
 this is the Landen transformation~\cite{Erde} on the complete
elliptic integral $\, K$. } take place~\cite{Erde}:
\begin{eqnarray}
&& _{2}F_{1}\Bigl( a,\,  b ; \, \,  2 b ; 16 w^2\Bigr)  
 \, = \, \, \,  \, \, \\
&&\qquad\qquad  \, \, \, \,  \, \,(1+ t^{1/2})^{2a} \cdot \, 
  _{2}F_{1}\Bigl( a, \, a -b  +1/2;\, b  +1/2;\, t \Bigr) \nonumber 
\end{eqnarray}

We have not been able to prove equality of the two kinds of formulas,
so, instead, we have resorted to comparison of their series expansions.
We have performed series expansions of our new integral representations
in the variable $\, t$ (see sec.(\ref{holo})), and found that they agree 
with the expansion of~(\ref{involved}).  In the next section, we will see that
they also agree with the coefficients of powers of 
$\, \lambda$, $\,  h_{2j}(N, \, N)(t)$ and 
$\,  h_{2j+1}(N, \, N)(t)$, in the series solutions
of~(\ref{jimbo-miwa})--(\ref{cpnorm}).

Our  new integral representations provide 
a ``closed enough'' representation of the $ \, \lambda$ 
coefficients of the various $ \, \lambda$ extensions 
 $ \, C(N, \, N; \lambda)$, the form factors. We
 will use the simplicity of these 
new integral representations in the next sections.

\section{Series solution of the sigma form of Painlev\'e VI}
\label{series1}

 In this section we study the series expansions of the diagonal
correlations $\, C(N, \, N)$ starting from 
(\ref{jimbo-miwa}), the sigma form of Painlev\'e VI.
 From order by order series analysis of the
solutions of the equation (\ref{jimbo-miwa}), 
we show, when $\,N$ is integer,
the existence of a one parameter $\, \lambda $ extension of  $\, C(N, \, N)$,
that actually identifies with the previous $\, \lambda $ extension
of a Toeplitz origin in~\cite{wu},  or of a
 Fredholm origin in~\cite{wu-mc-tr-ba-76}.  
 
We begin  by considering some remarkably simple
solutions of (\ref{jimbo-miwa}), which exist for all $N$ (not
necessarily integer). In particular, 
consider the $\, N$-dependent second order hypergeometric 
differential operator~\cite{PainleveFuchs} :
\begin{eqnarray}
\label{Ljm}
L_{h}\, = \, \, \, D_t^{2}\, 
+ \left( {{1} \over {t}} \, 
+\, {{1} \over {2 \, \left( t-1 \right)}}  \right)\cdot   D_t\,\, \,  
-{{1} \over {4}} \,{\frac {{N}^{2}}{{t}^{2}}}\,\, 
+{\frac {1}{16 \left( t-1 \right)^2}}\,
\end{eqnarray}
It has regular singularities at
$\, t=\, 0$, $t=1$ and $t=\infty$ with respectively
the critical exponents ($\pm N/2$), ($1/4, 1/4$) and ($1/4 \pm N/2$).
Denote by $\, h$  {\em any } solution of (\ref{Ljm}), 
and consider the $\, T \, > \, T_c$ 
expression (\ref{sigmahigh}) for $\, \sigma(t)$ :
\begin{eqnarray}
\label{sigmah}
\sigma(t)\,= \,\, t\, (t-1)\cdot  {d\ln h\over dt}\, -{1 \over 4}
\end{eqnarray}
The hypergeometric 
differential equation (\ref{Ljm}), when written  in $\, \sigma(t)$
given by (\ref{sigmah}), takes a ``Riccati'' form\footnote[2]{Note 
that, (\ref{Riccati}),  the Riccati form
of (\ref{Ljm}), is also covariant by the
 Kramers-Wannier duality (\ref{Kramers}). This is a quite
surprising result for a differential equation associated to 
$\, f^{(1)}(N, \, N)$, a form factor
one could think to be specific of the $\, T \, > \, T_c$ regime.} 
 :
\begin{eqnarray}
\label{Riccati}
&&16\,\,t \, (t-1) \cdot \sigma'  \, \,
 +16 \cdot \sigma^{2}\, \, -8\, \left( t-1 \right) 
\cdot \sigma \,  \, \,  \\
&& \qquad \qquad \qquad 
 - \, (2\,N\, -1)\, (2\,N+1)  \, (t-1)^{2} \, \, = \, \, \, \, 0 \nonumber 
\end{eqnarray}
For generic $ \, N$, it can be verified
 that $\, \sigma(t)$, given by (\ref{sigmah}),
is {\em actually a solution of} the sigma form
 of Painlev\'e (\ref{jimbo-miwa}), where $\, h$ is
{\em any linear combination} of the two 
solutions of (\ref{Ljm}) which, for generic $\, N$, read 
\begin{eqnarray}
\label{sol1}
f_{\pm}\, = \, \,\, \, 
t^{\pm N/2}  \cdot (1-t)^{1/4}\cdot  {_2}F_1( 1/2, 1/2\pm N;1\pm N; t)
\end{eqnarray}
and, for  integer $\, N$, are $ \, f_{+}$
and 
\begin{eqnarray}
\label{sol11}
t^{ N/2} \cdot  (1-t)^{1/4} \cdot {_2}F_1( [1/2, \, N \, +\, 1/2],\, [1], 1-t)
\end{eqnarray}
We recognize from (\ref{1}) that :
\begin{eqnarray}
f_+ \,=\,\,\, (1-t)^{1/4}\cdot {N! \over (1/2)_N} \cdot f^{(1)}_{N,N}
\end{eqnarray}

When $\,t\,\sim \, 0$ the leading behavior of
 $\, f_{+}$ is $\,f_{+}\sim t^{N/2}$
which, if we make the normalization
\begin{eqnarray}
\label{hyper}
h_N\,=\,\, \, {(1/2)_N\over N!} \cdot f_{+}\, \, \, 
=\, \,\, \,  (1-t)^{1/4}\cdot f^{(1)}_{N,N}
\end{eqnarray}
has the required behavior (\ref{cpnorm}) for the high-temperature
 two-point correlation function $\, C_{+}(N,N)$ as $\,t\,\sim\, 0$.
When the series expansion of $\, h_N$  is compared with
the series expansion of $\, C_{+}(N,N)$ \cite{PainleveFuchs}, we find that
\begin{eqnarray}
\label{boundary}
&&C_{+}(N,N)\, = \,\, \, (1-t)^{1/4}\cdot f^{(1)}_{N,N}\,   \\
&&\qquad \qquad +  {{(1/2)_N \cdot ((3/2)_N)^2 } \over {
 16 \cdot  \Gamma(N+2) \cdot  \Gamma(N+3)^2} }\cdot  t^{3N/2+2} \, 
+ \, \, \cdots \nonumber 
\end{eqnarray}
Thus, the first $\,N+1$ terms of $\,C_{+}(N,N)$, and
$\, \, (1-t)^{1/4}\, f^{(1)}_{N,N}$, 
coincide. The coefficient of $\, t^{3N/2+2}$ can be considered as an ``initial
condition'' needed to complete the characterization of the high-temperature
 two-point correlation function $\, C_{+}(N,N)$ seen as a solution of  
the Painlev\'e VI equation (\ref{jimbo-miwa}).

For $\, T<T_c$ one notes that 
\begin{eqnarray}
\sigma \, =\, \, t(t-1)\cdot {d\ln((1-t)^{1/4})\over dt}\, \, -{{t } \over {4}}
\end{eqnarray}
is a trivial solution of (\ref{jimbo-miwa}) and that 
\begin{eqnarray}
\label{boundary2}
\, C_{-}(N,N)\, =\, \, \, \, 
(1-t)^{1/4}\,\,\, 
  +{ {(1/2)_N \cdot (3/2)_N} \over {4\cdot  ((N+1)!)^2}}
 \cdot  t^{N+1}\, \,\,  +\,  \cdots
\end{eqnarray}
where the  coefficient
in front of  $\, t^{N+1}$ is the 
initial condition defining the two-point correlation function $\, C_{-}(N,N)$
in the low-temperature regime~\cite{PainleveFuchs,Ghoshalone}.

The relations (\ref{boundary}) and  (\ref{boundary2})
strongly suggest that, in order to analyze solutions of (\ref{jimbo-miwa}), 
we should  introduce the following form for the  low  temperature expansions
\begin{eqnarray}
\label{taulow}
C_{-}(N,N)\, = \, \, \, (1-t)^{1/4} \cdot   \Bigl(1 \,\, +\,
 \, \sum_{k=1}^{\infty}\, c_k \cdot t^k   \Bigr)
\end{eqnarray}
and, similarly, for the high temperature expansions :
\begin{eqnarray}
\label{tauhigh}
C_{+}(N,N)\,  =\, \,\,  (1-t)^{1/4}\cdot  f^{(1)}_{N,N}\, \, \,\,
 +(1-t)^{1/4} \cdot t^{N/2} \cdot \sum_{k=1}^{\infty}\,  d_k \cdot t^k 
\end{eqnarray}
These expansions are not the most general solutions of (\ref{jimbo-miwa})
because we have required that the
solutions have the correct behavior (\ref{cmnorm}) and (\ref{cpnorm})
at $\, t\,=0$. These forms yield a {\em one parameter} family of solutions. 

We consider first the low temperature regime and use the form (\ref{taulow})
 in  (\ref{jimbo-miwa}) to determine the $\, c_k$ 
coefficients recursively, order by order. 
When $\, N$ is not an integer this recursive procedure gives the unique
 solution $c_k=0$ for all $k.$ Thus the solution
 $\,(1-t)^{1/4}$ is the only solution of the form (\ref{taulow}).

However, when $\, N$ is an integer, we find that, while
$\, c_k =\,  0$ for $\, k \leq N$,  the equation which generically would
determine $c_{N+1}$ is automatically satisfied for all values of  
 $\, c_{N+1}$. The coefficient $\, c_{N+1}$ can be specified arbitrarily and
provides the second ``initial'' condition needed to specify a unique
solution of (\ref{jimbo-miwa}).

For all $\, k\, \geq N+2\, $ the order by order procedure uniquely determines
$\, c_k$ as a polynomial in  term of the free parameter 
$\,c_{N+1}$. More specifically, the term
$\,(c_{N+1})^n$ first appears in the coefficient $\,c_{n(N+n)}.$
Recalling (\ref{boundary2}) we see 
that if one writes the free parameter $\, c_{N+1}$ as
\begin{eqnarray}
c_{N+1}\,=\,\, \, \,   \, \lambda^2 \cdot 
{(1/2)_N \cdot  (3/2)_N \over 4 \cdot  ((N+1)!)^2}
\end{eqnarray}
the  order by order solution to (\ref{jimbo-miwa}) reads :
\begin{eqnarray}
\label{taulow2}
C_{-}(N,N;\lambda)
\,  =\, \,\,  \, 
 (1-t)^{1/4}\cdot \left(1\, + \sum_{n=1}^{\infty}\,
 \lambda^{2n}\cdot  h_{2n}(N)
\right)
\end{eqnarray}
where $\, h_{2n}(N)\sim t^{n(N+n)}\, $ for $\, t\,\sim\, 0$.
This $\, \lambda$ extension $\,  C_{-}(N,N;\lambda)$
reduces to the low-temperature two-point correlation function 
$\,C_{-}(N,N)$ when $\,\lambda=\,1$, and  to 
 $\,(1-t)^{1/4}\, $ for $\,\lambda=\, 0.$
Using  (\ref{taulow2}) in  (\ref{jimbo-miwa}) 
we have obtained the low-temperature series expansions 
of $\,  h_{2n}(N)$ for a large set
of values of the integer $\, N$, and  {\em found that these
 series expansions actually agree with the 
series expansions of} $(2n)$-multiple integrals $\, f^{(2n)}_{N,N}$ 
 defined in  sec. \ref{new}.

A similar order by order expansion can be carried out for the high
temperature case. The corresponding coefficients $\, d_{k}$ can be deduced
 recursively, order by order. When $\, N$ is not an integer the recursive
 procedure gives the unique solution $\, d_{k}=\,0\, $ for all $\, k$. 
For non integer $\,N$ we see that
 $\, h_N\,=\,(1-t)^{1/4}\cdot f^{(1)}_{N,N}\,$ 
is the unique solution of (\ref{jimbo-miwa}) of the form (\ref{tauhigh}).  

However, similar to the case for $\, T <T_c,$
we find that, when $N$ is an integer,
the  coefficients  $\, d_k$ are equal to zero for $k\leq N+1$,  and that the
coefficient $\,d_{N+2}$ is a free undetermined constant. The coefficients
$\, d_k$, for $\, k\, \geq N+3$, are polynomials
 in $\, d_{N+2}$. The term $\, (d_{N+2})^n$
 first appears in the coefficient $\, d_{n(N+n+1)}$.
 Thus, recalling (\ref{boundary}), we see that if we set 
\begin{eqnarray}
\label{singledouthigh}
d_{N+2} \,\, 
=\, \, \,\, \lambda^2 \cdot  {(1/2)_N \cdot  ((3/2)_N)^2  \over {
 16 \cdot \Gamma(N+2)\cdot  \Gamma(N+3)^2 }}    
\end{eqnarray}
the iterative solution to (\ref{jimbo-miwa}) may
 be written, in the high temperature
regime,  in the form
\begin{eqnarray}
\label{tauhigh2}
C_{+}(N,N;\lambda)\, =\,\,\,\,\,
 (1-t)^{1/4}\cdot \sum_{n=0}^{\infty}\, \lambda^{2n} \cdot  h_{2n+1}(N)
\end{eqnarray}
where
\begin{eqnarray}
h_1(N)\,\,=\,\,\, f^{(1)}_{N,N}
\end{eqnarray}
and where 
$\, h_{2n+1}(N)\, \sim \, t^{n(N+n+1)}$ for $\, t\, \sim\,  0$. This reduces to
 the high-temperature two-point correlation function
 $\, C_{+}(N,N)$ when $\,\lambda=\,1$ and to the hypergeometric
function $\, (1-t)^{1/4}\cdot  f^{(1)}_{N,N}$ for $\, \lambda =\, 0$.
Using (\ref{tauhigh2}) in (\ref{jimbo-miwa}) 
we have obtained the high-temperature series expansions 
of  $\,  h_{2j+1}(N)$'s for a large set
of values of the integer $\, N$, and  {\em found that these 
series expansions agree with}  $\,(2j+1)$-multiple 
integrals $\, f^{(2j+1)}_{N,N}$ defined in  sec.(\ref{new}).

We have also performed  low, and high, series expansions for the
$ \, \hat{C}^{j}(N, \, N)$ defined by equations (4.2) in~\cite{or-ni-gu-pe-01b}
(see also  (\ref{involved})), and we also {\em found that these series
identify with the one of }  $\,  h_{2j}(N)$ and 
$\,  h_{2j+1}(N)$
with the normalization :
\begin{eqnarray}
\label{hnormali}
h_{2j}(N) \, = \, \,  \hat{C}^{2j}(N, \, N), \qquad
h_{2j+1}(N) \, = \, \,  {{\hat{C}^{2j+1}(N, \, N)} \over {s}} \qquad
\end{eqnarray}
It would be most satisfying if these identities could be demonstrated
analytically.

All the previous results confirm that these various
 $\, \lambda$ extensions identify 
and actually verify (\ref{jimbo-miwa}).  The (log-derivative)
 of the $\, \lambda$ extensions
$\, C(N, \, N; \lambda)$ satisfy the same (sigma-form of ) Painlev\'e VI 
equation  (\ref{jimbo-miwa}) as the original diagonal
 spin-spin correlation, the boundary condition dependence
 coming from the  original diagonal 
spin-spin correlation boundary condition. Even if some might consider
that this result is not mathematically proved, it is clearly
an exact result of experimental mathematics, based on an accumulation of large
 computer formal calculations.

\section{Fuchsian linear differential equations for  $\, f_{N,N}^{(j)}(t)$ }
\label{holo}

In previous studies on the Ising 
susceptibility~\cite{ze-bo-ha-ma-04,ze-bo-ha-ma-05,ze-bo-ha-ma-05b,ze-bo-ha-ma-05c},
efficient programs
were developed which, starting from large series expansions of a
holonomic function, produce the 
linear ordinary  differential equation (in this case Fuchsian)
 satisfied by the function. In order for these
programs to be used to study the $\,f^{(j)}_{N,N}$'s  we need to efficiently
produce large (up to several thousand terms) series expansions in $t$ of the
$\,f^{(j)}_{N,N}$'s. We have done this by use of both the integral
representations (\ref{2n}), (\ref{2n-1}) and the representations 
of $\,f^{(j)}_{N,N}$ in terms of
theta functions of the {\em nome of elliptic functions}, 
presented in~\cite{or-ni-gu-pe-01b}.

We obtain the Fuchsian linear differential
 equations satisfied by the (diagonal)
form factors  $\, f^{(j)}_{N,N}$ for $\, j\, \leq 9$.
The analysis of these linear differential operators
 shows a remarkable Russian-doll structure
similar to the nesting of (the differential operators of)
 the ${\tilde \chi}^{(j)}$'s
 found
 in~\cite{ze-bo-ha-ma-04,ze-bo-ha-ma-05,ze-bo-ha-ma-05b,ze-bo-ha-ma-05c}. 
Specifically we find that the expressions
$ \, f^{(1)}_{N,N}$,  $ \, f^{(3)}_{N,N}$, $ \, f^{(5)}_{N,N}$,
 $ \, f^{(7)}_{N,N}$  are 
actually solutions of the linear ODE for $ \, f^{(9)}_{N,N}$, 
and that  $ \, f^{(0)}_{N,N}$, $ \, f^{(2)}_{N,N}$, 
$ \, f^{(4)}_{N,N}$, $ \, f^{(6)}_{N,N}$ are actually solutions 
of the ODE for $ \, f^{(8)}_{N,N}$.  
In addition, we find that all the linear differential operators
for the $\, f^{(j)}_{N,N}$'s have a direct sum decomposition in operators
equivalent to  symmetric powers of the differential operator corresponding to
$\, f^{(1)}_{N,N}$. Consequently,
 all the $\, f^{(j)}_{N,N}$'s  can also be
 written as {\em polynomials in terms of
 the complete elliptic integrals} $\, E$ and $\, K$.
The remainder of
this section is devoted to the presentation of these results.

\subsection{Fuchsian linear differential equations for $ \, f^{(2n+1)}_{N,N}$}
\label{Fuchsf2n-1}
The linear differential operator $\, F_9(N)$ which annihilates  
$ \, f^{(9)}_{N,N}$ has the
following factorized form :
\begin{eqnarray}
\label{F9}
F_9(N) \, = \, \, L_{10}(N) \cdot L_8(N) \cdot  L_6(N) \cdot
  L_4(N) \cdot  L_2(N)  
\end{eqnarray}
where the differential operators  $\, L_r(N)$ are of order $\, r$.
The first two read:
\begin{eqnarray}
 L_2(N)  \, = \,\,\,   \, Dt^2\,\, 
 + {\frac {2\,t-1}{ \left( t-1 \right) t}} \cdot Dt \, \, 
-{{1} \over {4 \, t}} +{{1} \over {4 \, (t-1) }}\, 
   - \,{\frac {{N}^{2}}{ 4 \, {t}^{2}}}
\end{eqnarray}
\begin{eqnarray}
 L_4(N)  \, = \,\,  \,\,  \, L_{4,0} \, \, - N^2\cdot L_{4,2} \,\,   
+{\frac {9}{16}}\,{\frac {{N}^{4}}{{t}^{4}}}
\end{eqnarray}
with:
\begin{eqnarray}
&&  L_{4,0}  \, = \, \, \,  Dt^4\, \, 
+10\,{\frac { (2\,t-1) }{ \left( t-1 \right) t}} \cdot Dt^3\, 
+\,{\frac { \left( 241\,{t}^{2}-241\,t+46 \right)}
{ 2\quad \,  \left( t-1 \right)^2\, t^2  }} \cdot Dt^2\nonumber \\
&& \quad +{\frac { \left( 2\,t -1\right) 
 \left( 122\,{t}^{2}-122\,t+9 \right) }
{  \left( t-1 \right)^{3}\, t^3  }}\cdot Dt \, \, 
+{\frac {81}{16}}\,{\frac { \left( 5\,t-1 \right) 
 \left( 5\,t-4 \right) }{{t}^{3} \left( t-1 \right) ^{3}}}
\end{eqnarray}
\begin{eqnarray}
  L_{4,2} \, = \, \, \,\,\,   {\frac {5}{2}}\,\,{\frac { Dt^2}{ t^2}} \, \, \, 
- \,{\frac { \left(23 \, - 32\,t \right) }
{ 2\,  \left( t-1 \right)\,  t^3  }} \cdot Dt \, \, \, 
-{\frac {9}{8}}\,{\frac {8-17\,t}{ \left(t-1 \right)\,  t^{4}}} 
\end{eqnarray}
The expressions (or forms) of  $\, L_6(N)$, $\, L_{8}(N)$ 
and $\,   L_{10}(N)$ are given in \ref{B}.  The
linear differential operators $F_{2n+1}(N)$, which annihilate
$f^{(2n+1)}_{N,N}$ for $n=0,\cdots, 3$, are such that: 
\begin{eqnarray}
\label{F7531}
&&F_7(N) \, = \, \,
 L_8(N) \cdot  L_6(N) \cdot  L_4(N) \cdot  L_2(N) \nonumber \\
&&F_5(N) \, = \, \,   L_6(N) \cdot  L_4(N) \cdot  L_2(N) \nonumber \\
&&F_3(N) \, = \, \,    L_4(N) \cdot  L_2(N)  \\
&&F_1(N) \, = \, \,    L_2(N)  \nonumber 
\end{eqnarray}
Thus we see that the differential operator for $\, f^{(2n-1)}_{N,N}$
{\em rightdivides} the differential operator for $\, f^{(2n+1)}_{N,N}$ for
  $\, n\, \leq 3$. We conjecture
 that this property holds for all values of $n$.
We thus have a  ``Russian-doll'' (telescopic) structure of 
these successive differential operators. 

\subsection{Fuchsian  linear differential equations for $ \, f^{(2n)}_{N,N}$}
\label{Fuchsianf2n}
The linear differential
 operator $\, F_8(N)$ (corresponding to $ \, f^{(8)}_{N,N}$)
 has the following factorized form: 
\begin{eqnarray}
\label{F8}
F_8(N) \, = \, \,\,
 L_9(N) \cdot  L_7(N) \cdot  L_5(N) \cdot  L_3(N) \cdot  L_1(N)  
\end{eqnarray}
where the linear differential operators  $\, L_r(N)$ are of order $\, r$.
The first two read:
\begin{eqnarray}
&& L_1(N)  \, = \, \, Dt,   \\ 
&& L_3(N)  \, = \, \, Dt^3 \, 
+4\,{\frac { \left( 2\,t-1 \right) }{t \left( t-1 \right) }} \cdot  Dt^2
+{\frac { \left(2 -15\,t +14\,{t}^{2} \right) }
{ \left( t-1 \right)^{2} \, t^{2}}}\cdot  Dt
\, 
\nonumber\\
&& \qquad \qquad \qquad + \,{\frac {8\,{t}^{2}-15\,t+5}
{ 2\,  \left( t-1 \right)^3 \,  t^2 }}\, \, \,
-\Bigl({\frac {{\it Dt}}{{t}^{2}}} \,
 +{{1} \over {t^3}}   \Bigr)\cdot  N^2   
\end{eqnarray}
The expressions (or forms) of  the linear differential operators
$\, L_5(N) $, $\,  L_7(N) $ and $\,  L_9(N) $
are given in \ref{B}. 

Similarly to (\ref{F7531}) there is a  Russian-doll (telescopic) structure of
 these successive linear  differential operators :  
\begin{eqnarray}
\label{F642}
&&F_6(N) \, = \, \, 
L_7(N) \cdot  L_5(N) \cdot  L_3(N) \cdot  L_1(N) \nonumber \\ 
&&F_4(N) \, = \, \, L_5(N) \cdot  L_3(N) \cdot  L_1(N) \nonumber  \\ 
&& F_2(N) \, = \, \,  L_3(N) \cdot  L_1(N) \nonumber  \\ 
&& F_0(N) \, = \, \, L_1(N) 
\end{eqnarray}
Again, we see that the linear differential operator for $\, f^{(2n-2)}_{N,N}$
{\em rightdivides} the linear differential operator for $\, f^{(2n)}_{N,N}$ for
  $\, n\, \leq 4$. We conjecture that
 this property holds for all values of $\, n$.

\subsection{Direct sum structure}
\label{beyond}
 
Not only do the linear  differential operators $L_{j}(N)$ have a factorized
Russian doll structure, but we have found that they also have a direct
sum decomposition when the integer $N$ is fixed. 
To illustrate this direct sum decomposition,
the corresponding linear differential operator for $\, f^{(3)}_{N,N}$ reads:
\begin{eqnarray}
\label{3Fdirect}
F_3(N) = \, \,    L_4(N) \cdot  L_2(N) 
\,  \, = \, \, \,    M_4(N) \oplus   L_2(N) 
\end{eqnarray}
where $\, L_2(N)$ is the linear differential operator for $\, f^{(1)}_{N,N}$
and the fourth order operator  $\, M_4(N) $
is displayed  in \ref{C} 
for successive values of $\, N$.
One remarks on these successive expressions
that the degree of each polynomial occurring in these linear 
 differential operators  $\, M_4(N)$ {\em grows linearly with} $\, N$. 

As a further example consider  $\, f^{(5)}(N,N)$,
where we find that the corresponding linear
 differential operator decomposes as:
\begin{eqnarray}
\label{5Fdirect}
F_5 \, \, = \, \, \,    L_6(N) \cdot  L_4(N) \cdot  L_2(N) 
 \, \, = \, \,   \,M_6(N)   \oplus M_4(N) \oplus   L_2(N) 
\end{eqnarray}
where $\, L_2(N)$ is the differential operator for $\, f^{(1)}_{N,N}$,
$\, M_4(N)$ is the previous fourth order differential operator,
and the sixth order operator  $\, M_6(N) $
has again coefficients whose degrees {\em grow 
with $ \, N$} for successive values of $ \, N$.
There is nothing specific to $\, f^{(3)}_{N,N}$ and $\, f^{(5)}_{N,N}$ :
similar results hold for all the $\, f^{(n)}_{N,N}$'s,
$\, n$ being even or odd. 

In contrast with the Russian-doll way of writing the differential
 operators for $\, f^{(n)}_{N,N}$, the direct sum structure,
as a consequence of this growing degree,   cannot be
 written for generic $N$ as operators with polynomials in front of
 the derivatives.
This ``non-closure'' of the direct sum structure will have some 
consequences when performing the scaling limit of
these differential operators (see sec.(\ref{scal}) below).  

\subsection{Equivalence of various $\, L_j(N) $'s and
 $\, M_j(N) $'s linear differential operators}
\label{equiv2}

We find that the symmetric square
\footnote[5]{The symmetric $j$-th power of
a second order linear differential operator
 having two solutions $\, f_1$ and  $\, f_2$
is the linear differential operator of order $j+1$, which has
 $\, f_1^{j}$, ... ,  $\, f_1^{j-k} f_2^{k}\, $,  ... ,
 $\, f_2^{j}$ as solutions. }
 of $\, L_2(N)$:
\begin{eqnarray}
&&{\rm Sym}^2(L_2(N)) \, \, = \, \,
 Dt^{3}
\, \, +3\,{\frac { \left(2\,t -1\right) }{ \left( t-1 \right) t}} \cdot Dt^2 
 +{\frac { \left( 1-7\,t+7\,{t}^{2} \right) }
{ \left( t-1 \right)^{2}{t}^{2}}}\cdot  Dt \nonumber  \\
&&\quad \quad -{{1} \over {2}} \,{\frac {1-2\,t}
 { \left( t-1 \right)^{2}{t}^{2}}} \, \, 
\, \, -{{N^2} \over {t}} \cdot Dt \, \,  
\, -{{N^2} \over {(t-1) \, t^2}}
\end{eqnarray}
and the linear differential operator $\, L_3(N) $ are equivalent :
\begin{eqnarray}
L_3(N)  \cdot U(N) \,\, = \, \,\,\,\, V(N) \cdot {\rm Sym}^2(L_2(N))
\end{eqnarray}
with the following intertwinners : 
\begin{eqnarray}
&&U(N) \, = \,\,
 \left(t -1 \right)\, t \cdot   Dt^{2}\, \, 
+ \left( 3\,t-1 \right) \cdot  Dt\, \, \, +1\, \, \, 
+{\frac { \left( 1-t \right) }{t}}\cdot {N}^{2}\, \, 
\\
&& V(N) \, = \,\, 
\left(t -1 \right)\, t \cdot  Dt^{2}\, \, 
+ \left( 11\,t-5 \right) \cdot Dt\, \, 
\, \, \ \nonumber \\
&& \qquad \qquad \quad +{\frac { \left( 5\,t-1 \right)
  \left( 5\,t-4 \right) }{ \left( t-1 \right)\, t }} \, \, 
\, \, -{\frac { \left( t-1 \right)}{t}} \cdot N^2 
\end{eqnarray}

Similarly, with the symmetric cube of $\, L_{2}(N)$, 
we have the equivalence
\begin{eqnarray}
 L_4(N) \cdot A(N) \,\,\,  = \,\,  \,\,\, B(N) \cdot {\rm Sym}^3(L_2(N)) 
\end{eqnarray}
with : 
\begin{eqnarray}
&&A(N)  \, =  \, \, ( t-1)\, t \cdot Dt^3\, \, 
+  {{7 }\over {2}} \, \left( 2\,t -1\right) \cdot  Dt^2 \, 
+ \,{\frac { \left( 41\,{t}^{2}-41\,t+6 \right)}
{ 4\,  \left( t-1 \right) t}} \cdot  Dt
 \nonumber \\
 &&\qquad  + {\frac {9}{8}}\,{\frac {2\,t-1}{ \left( t-1 \right)\,  t}} 
\,   -{\frac {9}{4}}\,{\frac { \left( t-1 \right) \cdot N^2 }{t}}\cdot  Dt\, 
-{\frac {9}{8}}\,{\frac { \left( 2\,t -1\right) }{{t}^{2}}}\cdot N^2 \\
&&B(N) \, =  \,  \, \, \left( t-1 \right)\, t \cdot Dt^{3} \, \,  \, 
 +{{23} \over {2}}\, \left( 2\,t-1 \right) \cdot  Dt^2 \,\, \nonumber \\
 && \quad\quad 
+{\frac {21}{4}}\,{\frac {
 \left( 6-29\,t+29\,{t}^{2} \right) }{ \left( t-1 \right) t}} \cdot Dt \, \, 
 +{\frac {9}{8}}\,{\frac { \left( 2\,t-1
 \right)  \left( 125\,{t}^{2}-125\,t+16 \right) }
{ \left( t-1 \right)^{2} \, t^2}}\nonumber \\
 && \quad \quad 
-{{9} \over {4}}\,{\frac { \left( t-1 \right) }{t}}
 \cdot N^2 \cdot   Dt\, \,  \,  \, 
-{\frac {9}{8}}\,{\frac { \left( 10\,t-9 \right) }{{t}^{2}}}   \cdot N^2 
\end{eqnarray}

More generally, all the $\, L_m(N)$'s are
 $(m-1)$-symmetric-power of  $\, L_2(N)$. 
As a consequence their solutions are $(m-1)$-homogeneous polynomial of the 
two hypergeometric  solutions of  $\, L_2(N)$.

Similarly, for the linear differential operators occurring in the direct sum, 
one easily verifies,  for every integer $\, N$, 
that, for instance,   the $\, M_4(N)$'s are equivalent to the 
 cubic-symmetric-power of  $\, L_2(N)$:
\begin{eqnarray}
\label{m4l2N}
       M_4(N) \cdot Q(N) \, \,  = \,  \,  \, S(N) \cdot {\rm Sym}^3(L_2(N)) 
\end{eqnarray}
where, for $\, N=\,\,  0,\, 1,\, 2$ :
\begin{eqnarray}
\label{forinstance}
&&Q(0) \,  \, = \, \, \left(t-1 \right)\, t \cdot   Dt\, \, 
 +t\, -{{1} \over {2}},\\
&&Q(1) \,  \,=   \, \, 2\, \left( t-1 \right)^{3}\, t^2 \cdot   Dt^{3}\,
+3\, \left( 3-7\,t+4\,{t}^{2} \right)  \left( t-1 \right)\,  t \cdot   Dt^{2}
\nonumber \\
&&\quad + \left( 12\,{t}^{3}-28\,{t}^{2} +{\frac {41}{2}}\,t\, 
-{{9} \over {2}} \right) \cdot Dt\,\, \, 
+{{3} \over {4}}\,{\frac{2\,{t}^{2} -2\,t+1}{t}},\\
&&Q(2) \,  \, =  \, \,
{{1} \over {3}}\, \left( t-1 \right)^{3}\, 
 \left( 3+8\,t+3\,{t}^{2} \right)\, t \cdot  Dt^{3}\,  \nonumber \\
&&\quad +{{1} \over {2}}\,
 \left( 15-t-35\,{t}^{2}+15\,{t}^{3}+6\,{t}^{4} \right)  
\left( t -1 \right)\cdot  Dt^{2} \nonumber \\
&&\quad -{{1} \over {24}}\,{\frac { \left( 18\,{t}^{5}
-12\,{t}^{4}-97\,{t}^{3}+577\,{t}^{2}-738\,t+252
 \right) }{t}  } \cdot  Dt\,  \nonumber \\
&&\quad
-{{1} \over {16}}\,{\frac {12\,{t}^{5}+14\,{t}^{4}
-260\,{t}^{3}+497\,{t}^{2}-314\,t+24}{{t}^{2}}} 
\end{eqnarray}

As a further example, one can verify,
for every value of the integer $\, N$, 
that  the sixth order operator  $\, M_6(N) $ is equivalent
to the fifth symmetric power of $ \, L_2(N)$.
The solutions of the linear differential operators
$ \, M_m(N)$ are also $(m-1)$-homogeneous polynomials of the
two hypergeometric  solutions of  $\, L_2(N)$.

As a consequence of this  direct sum decomposition,
 the solutions  $\, f^{(n)}(N,\, N)$
are (non-homogeneous) {\em polynomials of  the 
two hypergeometric  solutions of}  $\, L_2(N)$ or, 
 equivalently,  $\, f^{(1)}_{N,N}$ (or $\,  h_N$ see (\ref{hyper})) and its 
first derivative.  The second order linear differential operator $\,  L_2(N)$
is equivalent~\cite{PainleveFuchs} to the 
second order  differential operator $\, L_E$
\begin{eqnarray}
\label{LE}
L _E \, = \, \, \, \, 4\,t \cdot  Dt^{2}\,\, + 4\, Dt\, - {{1} \over {t-1}}
\end{eqnarray}
 corresponding to the complete elliptic integral $\, E$. As a consequence
of the previously described direct sum decomposition, the $\, f^{(n)}_{N,N}$'s
can also be written as {\em polynomial expressions of  the
 complete elliptic integral} $\, E$ and its first derivative
 $\, E'$, or alternatively,  $\, E$ and
  {\em the complete elliptic integral} $\, K$.

\vskip .1cm

Let us just give here a set of miscellaneous 
examples. For $\, f^{(2)}_{N,N}$, one has:
\begin{eqnarray}
\label{miscell}
&& 2\, \,  f^{(2)}_{0,0} \,  =\, \, 
 \left(K\, -E \right)\cdot  K  \nonumber \\
&& 2\, \, f^{(2)}_{1,1} \,  = \, \, 
1\, -3\,K\, E\, - \left(t-2 \right)\cdot  {K}^{2} \nonumber \\
&& 6\, \, t\cdot  f^{(2)}_{2,2} \,  =\, \, 
6\,t \, - \left( 2+6\,{t}^{2}-11\,t \right) \cdot {K}^{2}\,  \nonumber \\
&&\quad \quad - \left( 15\,t -4 \right)\cdot K\, E\,\,\,
 -2\, \left( 1+t \right)\cdot  {E}^{2}  \\
&& 90 t^2 f^{(2)}_{3,3} \,  =\, \, 135\,{t}^{2} 
- \left( 137\,{t}^{3}-242\,{t}^{2}+52\,t+8 \right)\cdot  {K}^{2} \nonumber \\
&& \quad + \left( 8\,
{t}^{3}-319\,{t}^{2}+112\,t+16 \right)\cdot  KE-4\, \left( 1+t \right) 
 \left( 2\,{t}^{2}+13\,t+2 \right) {E}^{2}  \nonumber \\
&& 3150\, \, t^3 \cdot  f^{(2)}_{4,4} \,  =\,\, \, 6300\,{t}^{3} \nonumber \\
&&\quad - \left( 32\,{t}^{5}+2552\,{t}^{2} +128+6440\,{t}^{4}-11191\,{t}^{3}\, 
+464\,t \right)\cdot {K}^{2} \nonumber \\
&&\quad + \left( 128\,{t}^{5}+5648\,{t}^{2}-14519\,{t}^{3}\, 
+1056\,t +576\,{t}^{4}+256 \right)\cdot  E\, K \nonumber \\
&&\quad -8\, \left( 1+t \right)\,
  \left(16\,{t}^{4}+58\,{t}^{3}+333\,{t}^{2}\,
 +58\,t+16 \right)\cdot  {E}^{2}  \nonumber 
\end{eqnarray}
where $\, E$ and $\, K$ are given by (\ref{EK}). 
Other examples are given in \ref{AA}.

\vskip .1cm

{\bf Remark:} All these remarkable structures {\em are not
 restricted to diagonal two points correlation functions}.
Actually one can calculate various 
$\, j$-particle contributions $\, f^{(j)}_{M,N}$
of  the {\em off-diagonal} two point correlation functions, 
and verify, again, that they are {\em also polynomial 
expressions of the complete elliptic integrals}
$\, E$ and $\, K$. For instance for $\, T\, >\, T_c$ :
\begin{eqnarray}
\label{nice}
&&C^{(2)}(0, \, 1) \, = \, \, 
{{3} \over {8}}\,\,  \,
 - {{1} \over {4}}\, \left( 1+{s}^{2} \right)\,  K\,  \\
&& \quad\quad \quad\quad\quad\quad
-{{1} \over {2}}\,E\, K\, \, 
\, -{{1} \over {8}}\, \left( {s}^{2}-3 \right) 
 \left(1+{s}^{2} \right)\cdot  K^{2} \nonumber 
\end{eqnarray}
where $\, s \, = \, \sinh(2\, K)$.
Other miscellaneous examples of such off-diagonal
  $\, j$-particle contributions
are displayed in \ref{D}.

\subsection{The elliptic representation of Painlev\'e VI}
\label{elliptic}

The results we have underlined in this section, namely the unexpectedly
simple and remarkable polynomial expressions 
for the form factors $\, f^{(j)}_{N,N}$,
 correspond to the fact that the associated linear differential operators
are direct sums of operators equivalent to
 symmetric powers of the second order differential operator $\, L_E$.
We already encountered this central key role played
 by the linear differential operator $\, L_E$,
or the hypergeometric second order differential operator (\ref{Ljm}),
in our previous holonomic analysis of the two-point correlation
functions of the Ising model~\cite{PainleveFuchs}.
In order to understand the key role played by $\, L_E$,
or equivalently operator (\ref{Ljm}), it is worth
 recalling (see~\cite{man}, or for a review~\cite{Guzzetti}) the so-called 
``{\em elliptic representation}'' of
 Painlev\'e VI. This  elliptic representation
of Painlev\'e VI amounts to seeing Painlev\'e VI as a 
``deformation'' (see equation (33) in~\cite{Guzzetti})  of the 
hypergeometric linear differential equation 
 associated with the linear differential operator :
\begin{eqnarray}
{\cal L} \, = \, \, \, \, (1-t) \,  t \cdot Dt^2 \,  \,
 + (1-2\, t) \cdot Dt \, \,  -{{1} \over {4}} 
\end{eqnarray}
One easily verifies that this linear differential operator is actually 
equivalent (in the sense of the equivalence
of differential operators) with $\, L_E$, or equivalently (\ref{Ljm}).
This deep relation between {\em elliptic curves and Painlev\'e VI}
explains the occurrence of Painlev\'e VI on the Ising model,
and on other lattice Yang-Baxter integrable models
which are canonically parametrized in term of elliptic functions 
(like the eight-vertex Baxter model, the RSOS models, 
see for instance~\cite{Manga}). 
We will see, in sec.(\ref{algebr}), another example of this deep connection 
between the transcendent solutions
of Painlev\'e VI and the theory of elliptic functions, modular curves
 and quasi-modular functions. 

\section{The scaling of  $f^{(j)}_{N,N}$}
\label{scal}

The scaling of  the $ \, f^{(n)}_{N,N}$'s amounts, on the functions,
and on the corresponding differential operators, to
 taking the limit $\, N \, \rightarrow \, \infty$
and $\, t \, \rightarrow \ 1$, keeping the limit 
$ \, x \, = \, \, N\cdot (1-t)$ finite, or in other words, to
 performing the change of variables  $\, t=1-x/N$,
keeping only the leading term in $\, N$. 
Performing these straightforward calculations, the linear
differential operators in $\, t$
for the $ \, f^{(n)}_{N,N}$'s where $\, N$ was a parameter,
 become linear differential operators in the only 
scaling variable $\, x$. 

Calling $\, F^{scal}_j$ the scaling limit of the operator $\, F_{j}(N)$ we
find for $j$ even that
\begin{eqnarray}
\label{f6scal}
&&F_{8}^{scal}  \, = \, \,
 L_9^{scal} \cdot  L_7^{scal} \cdot  L_5^{scal}
 \cdot  L_3^{scal} \cdot  L_1^{scal}  \nonumber \\
&&F_{6}^{scal}  \, = \, \, L_7^{scal} \cdot  L_5^{scal} 
\cdot  L_3^{scal} \cdot  L_1^{scal}  \nonumber \\
&&F_{4}^{scal}  \, = \, \,  L_5^{scal} \cdot
  L_3^{scal} \cdot  L_1^{scal}  \nonumber  \\
&&F_{2}^{scal}  \, = \, \,  L_3^{scal} \cdot  L_1^{scal}  \nonumber \\
&&F_{0}^{scal}  \, = \, \,  L_1^{scal}  
\end{eqnarray}
where :
\begin{eqnarray}
&&L_5^{scal}  \, = \, \, 2\,{x}^{5}{{\it Dx}}^{5}
+10\,{x}^{4}{{\it Dx}}^{4}-2\,{x}^{3} \left( 7+5\,{x}^{2} \right) 
{{\it Dx}}^{3} \nonumber \\
&&\quad  +2\, \left( -16+13\,{x}^{2} \right)\, {x}^{2} {{\it Dx}}^{2}
 +2\, \left( 5-12\,{x}^{2}+4\,{x}^{4} \right)\, x\,  {\it Dx} \nonumber \\
&&\quad  -10 +8\,{x}^{2} -24\,{x}^{4},\nonumber \\
&&L_3^{scal}  \, = \, \,2\,{x}^{3}{{\it Dx}}^{3}\, +8\,{x}^{2}{{\it Dx}}^{2}
-2\, \left( x-1 \right)  \left( x+1 \right)\, x \, {\it Dx} \, 
-2, \nonumber \\
&&L_1^{scal}  \, = \, \,Dx   
\end{eqnarray}
and $\, L_9^{scal}$, $\, L_7^{scal}$ are given in \ref{E}.

Similarly, for $j$ odd, we have
\begin{eqnarray}
\label{f7scal}
&&F_{9}^{scal}  \, = \, \, 
L_{10}^{scal} \cdot L_8^{scal} \cdot  L_6^{scal}
 \cdot  L_4^{scal} \cdot  L_2^{scal} \nonumber  \\
&&F_{7}^{scal}  \, = \, \, L_8^{scal} \cdot
  L_6^{scal} \cdot  L_4^{scal} \cdot  L_2^{scal} \nonumber  \\
&&F_{5}^{scal}  \, = \, \,L_6^{scal} \cdot
  L_4^{scal} \cdot  L_2^{scal}   \nonumber \\
&&F_{3}^{scal}  \, = \, \,  L_4^{scal} \cdot  L_2^{scal}  \nonumber \\
&&F_{1}^{scal}  \, = \, \,   L_2^{scal} 
\end{eqnarray}
where  
\begin{eqnarray}
&&L_4^{scal}  \, = \, \,16\,x^4\,  Dx^4\, +96\,x^3\,  Dx^3
+ 40\, \left( 2-x^2 \right)\,  x^2\,  Dx^2\nonumber \\
&&\quad \quad \quad +8\, \left( {x}^{2}-2 \right) \,x \,  Dx\,\, 
+9\,x^{4} -8\,{x}^{2}+16,\nonumber  \\
&&L_2^{scal}  \, = \, \,4\,x\,  Dx^{2}\,\, +4\, Dx \, \, -x 
\end{eqnarray}
and $\, L_{10}^{scal} $,  $\, L_8^{scal} $,  $\, L_6^{scal} $ 
are given in \ref{E}.

Thus, we see that the scaled operators $\, F_j^{scal}$ have a ``Russian doll''
structure inherited from the lattice operators $F_j(N)$.

Consider the linear differential operator corresponding to the modified
Bessel function $Bessel(n,x/2)$ for $\, n=0$, namely
\begin{eqnarray}
\label{modBessel}
 B\, = \, {{\it Dx}}^{2}\,\, +{\frac {{\it Dx}}{x}}\,\, - {{1} \over {4}}
\end{eqnarray}
We recognize, in this linear differential
 operator, the exact identification with
the scaled differential operator $F_1^{scal}=L_2^{scal}$. 
We find  that the symmetric square of the linear
 differential operator $\, B$, and the scaled 
operator $\, L_3^{scal} $ {\em are equivalent} : 
\begin{eqnarray}
&& L_3^{scal}  \cdot ( x {{\it Dx}}^{2}\, +2\,{\it Dx}\, -x) \,   = \,    \\
&& \qquad \quad  (2\,{x}^{4}{{\it Dx}}^{2}+12\,{x}^{3}{\it Dx}\, 
-2\,{x}^{4}+8\,{x}^{2}) \cdot     {\rm Sym}^2(B)  \nonumber
\end{eqnarray}
Similarly, the symmetric third power of the
 linear differential operator $\, B$,
 and  the scaled operator $\, L_4^{scal} $ are equivalent, 
and, more generally, the symmetric $j$-th power  of
 (\ref{modBessel}) and the scaled operator
 $\, L_{j+1}^{scal} $ {\em are equivalent} :
\begin{eqnarray}
L_{j+1}^{scal} \, \, \simeq \,\,    {\rm Sym}^j(B) 
\end{eqnarray}

Recall that the differential operators $F_j(N)$, corresponding
 to the form factors $f^{(j)}_{N,N}$, can 
be written as direct sums {\em only when} the integer $\, N$ is fixed.
At the scaling limit, this feature dissappears 
in the scaled differential operators
$\, F_j^{scal}$ which have no direct sums. 
Therefore while the scaling limit
 preserves the Russian-doll (telescopic) 
structure (see  (\ref{F7531}), 
 (\ref{f7scal})) and also preserves 
 the fact that the various operators in this Russian-doll (telescopic) 
structure are equivalent to symmetric  powers of an operator (\ref{modBessel})
which replaces the operator $\, L_E$, 
the {\em direct sum structure is lost}. 
As a consequence the scaling of the   $\, f^{(j)}_{N,N}$'s 
{\em cannot be seen as simple polynomials of modified Bessel functions}. 

There is one exception that concerns $\, f^{(2)}_{N,N}$. Its scaled linear
differential operator $\, F_{2}^{scal} \, $, has the non shared property of
being equivalent to the {\em direct sum} of   $\, Dx$
 with  the symmetric square of  (\ref{modBessel}), namely: 
\begin{eqnarray}
F_{2}^{scal} \, \, = \,\,  \,
 L_1^{scal} \oplus  L_3^{scal} \,\,  \simeq \,\,  \, Dx \oplus 
{\rm Sym}^2(B) 
\end{eqnarray}
From this equivalence, one immediately deduces the expression of the scaling
of the $\, f^{(2)}_{N,N}$ as {\em quadratic expression of the modified 
Bessel functions} of $\, x/2$ which actually identifies
with formula (2.31b)-(3.151) in~\cite{wu-mc-tr-ba-76}.

\vskip .1cm

 The occurrence of modified Bessel functions, emerging from a 
confluence of two singularities of the
 complete elliptic integrals $\, E$ and  $\, K$, or from
the hypergeometric function $ \, _{2}F_1$, should not be considered
as a surprise if one recalls the following limit of 
the hypergeometric function $ \, _{2}F_1$ yielding 
confluent hypergeometric functions
$ \, _{1}F_1$. These confluent hypergeometric functions,
$ \, _{1}F_1$, are nothing but modified Bessel functions~\cite{Erde}:
\begin{eqnarray}
&& _{1}F_1(a, \, b; \, z) \, \quad
\rightarrow \quad  _{2}F_1(a, \,p, \, b; \, {{z} \over {p}})
\quad \quad \hbox{when:}
 \quad p \,  \quad
\rightarrow \quad \infty \nonumber \\
&& I(\nu, \, z) \, = \, \, \, \,
 {{ z^{\nu} } \over {2^{\nu} \, e^z \, \Gamma(\nu+1) }} \cdot \,
 _{1}F_1(\nu \, +{{1} \over {2}}, \, 2\, \nu\, +1; \, 2\, z)
\end{eqnarray}

\vskip .1cm

{\bf Remark:} It was shown, in sec.(\ref{holo}), as a consequence of 
 the decomposition of their differential operators in direct sums of operators
equivalent to symmetric
 powers of $\, L_E$, that the  functions $f^{(n)}_{N,N}$ are 
polynomial expressions of $\, E$ and $\, K$ 
functions. Therefore their singularities 
are only the three {\em regular} points 
$\, t =0$, $\, t =1$ and $\, t =\infty$.
The scaling limit
 ($\, t=\, 1-x/N$, $t \rightarrow 1$, $N \rightarrow \infty$)
 corresponds to the {\em confluence}
 of the two regular points $\, t =0$ and $\, t =\, \infty$, yielding
the, now, {\em irregular}  point $\, x=\infty$.
The occurrence of irregular points with their Stokes phenomenon,
 and, especially, the {\em loss of a remarkable  direct sum structure}, 
shows that the scaling limit is a quite non-trivial limit.

Contrary to the common wisdom, the scaling limit does not correspond 
to more ``fundamental'' symmetries and structures (more universal ...) : 
this limit {\em actually destroys the remarkable
 structures and symmetries of the
 lattice models}\footnote[3]{These kind of results should  not be 
a surprise for the people working on integrable lattice models, 
or on Painlev\'e equations~\cite{Sakai1,Sakai2}.}.

\section{Algebraic solutions of PVI for
 $\lambda=\cos(\pi m/n)$ and modular curves}
\label{algebr}

The function $\, C(N,N;\lambda)$ is such that its log-derivative
is actually a solution of the sigma form of Painlev\'e VI : it is 
a {\em transcendent}
 function ``par excellence''. However, the unexpectedly simple
expressions for these form factors $\, f^{(j)}_{N,N}$,
 strongly suggest to try to
resum the infinite sums (\ref{formm1}), and (\ref{formp1}), 
of  form factors $\, f^{(j)}_{N,N}$, corresponding
to the function $\, C(N,N;\lambda)$, 
and see if these transcendent functions could be ``less complex'' than
one can imagine
at first sight, at least for a set of ``singled-out''
 values of $\, \lambda$. For instance,
 are there any values of $\lambda  \neq 1$ 
which share, with $\, \lambda\, =1$, the
property that $\, C(N,N;\lambda)$ satisfies 
a Fuchsian linear differential equation ?

Actually, introducing, instead of the modulus  $\, k$ 
of elliptic functions (for $\, T \, > \, T_c$,  $\, k\, = s^2$), 
or  the $\, s$ and $\, t$ variables,
 the {\em nome} of the elliptic functions
 (see relations  (5.7)-(5.11) in~\cite{or-ni-gu-pe-01b}),
we have been able to perform such a resummation,
 getting, {\em for arbitrary} $\, \lambda$, nice closed 
expressions for the  $\, C(N,N;\lambda)$ for the first 
values of $\, N$, ($N\, = \, 0, \, 1\, , \, 2\, \cdots $),
as sums of ratios of theta functions (and their derivatives), corresponding
 to  {\em Eisenstein series}, or {\em quasi-modular
forms}. These results will be displayed
 in forthcoming publications. The simplest 
example corresponds to $\, N\,= 0$  where 
 $\, C_{-}(N,N;\lambda)$ 
is just the ratio of two 
Jacobi  $\, \theta_3$ functions :
\begin{eqnarray}
\label{closed}
C_{-}(0,0;\lambda)\, = \, \, \, 
{{ \theta_3(u,q)} \over {\theta_3(0,q)}},
  \quad \quad  \hbox{where:} 
\quad\quad\quad \quad \lambda\,=\,\,\cos(u)  
 \quad
\end{eqnarray}
All these results strongly suggest to focus
 on $\, u\, = \, \pi \, m/n$ ($m$ and $\, n$ integers)
yielding for the possible choice of ``singled-out'' values of $\, \lambda$ :
\begin{eqnarray}
\label{special}
\lambda\,=\,\, \cos(\pi m/n)
\end{eqnarray}
Actually  these special values (\ref{special}) of
$\, \lambda\, $ already occurred in a study of $\, N=2$
 supersymmetric field theories~\cite{cv}
in a similar series construction of solutions of the Painlev{\'e} V
(or Painlev{\'e} III for a ratio of functions)
equation for the scaling limit of the Ising model~\cite{wu-mc-tr-ba-76}.

We have begun to investigate this situation. When 
$\, n=\, 3,\, \cdots,\, 20$ (and all the possible values
of $\, m$, but a set of first successive values of $\, N$), we have found
that $\, C_{\pm}(N,N;\lambda)$ do indeed satisfy
 Fuchsian linear differential 
equations but, unlike the equations found in~\cite{PainleveFuchs} for
$\lambda\, =\, 1$, the order of
the Fuchsian linear differential equations
 {\em depends only on} $n$ {\em and not on} $\, N$. 

As examples of these Fuchsian linear differential 
equations, we found, for instance, that 
$\, C_{-}(N,N;\cos(\pi/4))$, for $\, N\, =\, 0,\, 1,\, 2$, are annihilated,
respectively, by
\begin{eqnarray}
\label{1over4}
&&L^{[1/4]}_0\,=\,\,\, (t-1)^2\,t \cdot Dt^2\,\,
+{3\over 8}(t-1)(3t-2)\cdot Dt\,\,\,\,
-{15\, t\over  256}\,\,+{3\over 32} \nonumber \\
&&L^{[1/4]}_1\,=\,(t-1)^2\,t\cdot Dt^2\,+(t-1)(5t-2) \cdot Dt\,\,
 -{7\, t \over  256}\,+{1\over 16}\nonumber\\
&&L^{[1/4]}_2\,=\, \, (t-8)(t-1)^2\,t \cdot Dt^2\,\, +(t-1)(t^2-2t+16)\cdot Dt
\nonumber\\
&& \quad\quad \quad \quad \quad +{209\, t^2\over 256}\,
 -{25 t\over 16}\,+{1\over 2} 
\end{eqnarray}
and that $\, C_{-}(0,0;\cos(\pi/3))$ is annihilated by :
\begin{eqnarray}
\label{1over3}
&&L^{[1/3]}_0\,=\,\,(t-1)^3\,t^3 \cdot Dt^4\,\,\,
 +{11 \over  3} \, \,(2t-1)\, (t-1)^2\,t^2\cdot Dt^3 \nonumber\\
&&\quad \quad +{7\over 27}\, (43t^2-43t+4) (t-1)\, t\cdot Dt^2 \\
&&\quad \quad 
+{7\over 1458}\, (2t-1)(247t^2-247t-80) \cdot Dt \,\,
\, \,+{35 \over  486}\nonumber 
\end{eqnarray}
These linear differential operators are 
of a {\em quite different nature} from the one depicted in sec.(\ref{holo})
which can be decomposed in direct sums of (operators equivalent to) 
symmetric powers of $\, L_E$.
In contrast with the direct sum decomposition
we have underlined  previously, these linear differential operators
are {\em irreducible}. 
However we do expect from sec.(\ref{elliptic}) a
 connection with elliptic curves. 
Actually, instead of a connection through the second
 order differential operator $\, L_E$, or
the hypergeometric second order linear differential
 operator (\ref{Ljm}), we have an even more
striking link with the theory of elliptic curves. These
 solutions $\, C(N,N;\lambda)$
are actually {\em algebraic solutions}
 of Painlev\'e VI, associated 
with {\em modular curves}\footnote[3]{The occurrence of
modular curves is pretty clear for $\, N=0$
 from (\ref{closed}), from the analysis of its 
invariance group,  subgroup of the modular group.}.
We found for $\, n\, \leq\,  8$ these singled-out 
Fuchsian linear differential equations, corresponding
to algebraic solutions of Painlev\'e VI, 
and beyond, directly these modular curves for larger
 values of $\, n$ for which we do not have the  
Fuchsian linear differential equations yet.

We first obtained these modular curves as polynomial
 relations $\, P(\sigma, \, t)=\, 0$, between 
$\, \sigma$ and $\,  \, t$, and we then found, in a 
second step, the polynomial relations
  $\, P(\tau, \, t)=\, 0$, between  $\,  \tau\, =
 \, C_{\pm}(N,N;\cos(\pi \, m/n))$
and $\, t$. For instance, one finds that $\,\tau \,
 = \, C_{-}(0,0;\cos (\pi/3))\, $ is solution of 
a  genus {\em one} algebraic curve :
\begin{eqnarray}
\label{curve13}
16\,\tau^{12}\,-16\,\tau^9\,\,
 -8  (t-1) t \cdot \tau^3\,  +t \cdot (1-t)\,\,=\,\,\,0
\end{eqnarray}
or that  $\,\, \tau \, = \,C_{-}(N,N;\cos(\pi/4))\,$ is solution of 
genus {\em three} algebraic curve, for instance, for $\, N=\, 0$,
\begin{eqnarray}
\label{curve16}
16\,\tau^{16}\,+16\,(t-1)\cdot  \tau^8\,   +t^2 \cdot (t-1)\,\,=\,\,\,\,0
\end{eqnarray}
the  corresponding solutions being quite simple
algebraic expressions :
\begin{eqnarray}
&&C_{-}(0,0;\cos(\pi/4)) \, = \,\,
2^{-1/4}(1-t)^{1/16}[1+(1-t)^{1/2}]^{1/4} \label{c1}  \\
&&C_{-}(1,1;\cos(\pi/4)) \, = \,\,
2^{-3/4}(1-t)^{1/16}[1+(1-t)^{1/2}]^{3/4} \label{c2}  \\
&&C_{-}(2,2;\cos(\pi/4)) \, = \,\,  \\
&&\qquad\qquad =\,  2^{-5/4}(1-t)^{1/16}[1+(1-t)^{1/2}]^{5/4}\label{c3}
[5-(1-t)^{1/2}]/4\nonumber 
\label{curve14}
\end{eqnarray} 

We give in Table 1, when available, the order of the Fuchsian
linear differential equation for 
$\, \lambda\, =\, \cos(\pi /n)$, the degree and genus of the
corresponding algebraic curve  $\, P(C_{-}(0,0;\lambda), \, t)=0$, and the
degree and  genus of the algebraic 
$\sigma$-curve $\, P(\sigma(0,0;\lambda), \, t)=0$.

\begin{center}
\begin{table}[h!]
\label{table1}
\begin{tabular}{|l|l|l|l|l|l|l|l|l|l|l|l|l|l|l|} \hline
$n$ & 3 & 4 & 5 & 6 & 7 & 8 & 9 & 10 & 11 & 12 & 14 & 16 & 18 & 20  \\
 \hline 
ODE order  & 4  & 2  & 12 & 4 & 24 & 8 &  &  &    & &  & & &  \\
 \hline 
$\tau$-degree  & 12  & 16  & 60 & 48  & 168 & 128 &  & 240   &  & &  &  & &  \\\hline
$\tau$-genus  & 1  & 3  &  13 & 13 &  & 41 &  &  &   & &  & & &  \\
 \hline 
$\sigma$-degree  & 4  & 2  & 12 & 4 & 24 & 8 & 36 & 12 & 60 & 16  & 24 & 32 & 36  & 48 \\
 \hline 
$\sigma$-genus  & 0  & 0  & 1 & 0 & 4 & 0 &  & 1 &  & 1 &  & &  &   \\
 \hline 
\end{tabular}
\caption{Order of the linear ODE, as well as
 degree and genus of the corresponding
modular curve in $\, \tau \, = C_{\pm}(0,0;\lambda)$ 
 for  $\, \lambda=\, \cos(\pi /n)$,
when available. The corresponding degree and genus of the modular curve in
$\sigma(0,0;\lambda)$, when available.}
\end{table}
\end{center}

We found the following results on the polynomial
 relations  $\, P(\tau, \, t)=\, 0$, between
  $\,  \tau\, = \, C_{-}(N,N;\cos(\pi \, m/n))$ and $\, t$. These polynomials
are {\em actually polynomials of the variable}
  $\, \rho\, = \, \tau^n\, $ for $\, n$ odd 
and  of the variable  $\, \rho\, = \, \tau^{2\, n}\, $ for $\, n$ even. 
This property is related to the  invariance of the variable
 $\, \rho$  under a subgroup of the modular
group\footnote[5]{See in particular Barth and Michel~\cite{Barth} for 
further details on the $\, X_{00}(n, \, 2)$ modular 
curves and the characterization of
the genus of  modular curves  from subgroups of 
$\, SL(2, \, Z)$. We will study  $\, C_{\pm}(N,N;\lambda)$
 from this modular subgroup point of view elsewhere. }. Let us denote
 $\, Q(\rho, \, t)=\, 0$ the  polynomial relation between $\, \rho$ and $\, t$.
We also found that the degree of the polynomial  $\, Q$ in $\, \rho$
actually identifies with the degree in $\, \sigma$ 
of the polynomial  $\, P(\sigma, \, t)=\, 0$.
 Thus, the $\tau$-degree in Table 1
can be seen to be the $\sigma$-degree multiplied by $\, n$ for $\, n$ odd,
 and by $\, 2\, n$ for $\, n$ even. The order of the 
Fuchsian linear differential equations for $\, C_{-}(N,N;\lambda)$
identifies with that degree in $\, \sigma$. 
We finally found that the genus of the modular curve $\, P(\sigma, \, t)=\, 0$
identifies with the genus of the $\, \tau^n$ (resp. $\, \tau^{2\,n}$)-modular
 curve  $\, Q(\rho, \, t)=\, 0$ : the genus corresponding to 
 $\, C_{-}(0,0;\cos(\pi/3))^3 $,  $\, C_{-}(0,0;\cos(\pi/5))^5 $ 
are respectively $\, 0$, $\, 1$,  the genus corresponding to
  $\, C_{-}(0,0;\cos(\pi/4))^8 $,  $\, C_{-}(0,0;\cos(\pi/6))^{12} $,
  $\, C_{-}(0,0;\cos(\pi/8))^{16} $ are  $\, 0$ but the genus for
  $\, C_{-}(0,0;\cos(\pi/10))^{20} $ is $\, 1$. In contrast the 
 genus corresponding to   $\, C_{-}(0,0;\cos(\pi/6))^{6} $,
  $\, C_{-}(0,0;\cos(\pi/8))^{8} $ are  $\, 1$, and the  genus for 
 $\, C_{-}(0,0;\cos(\pi/10))^{10} $ is $\, 5$. 

For $\, N\, =\,0$, and only in this case, a 
large set of these algebraic curves  
(for instance (\ref{curve13}) or the modular curve 
for $\, n\, = 7$ in the previous table) 
are invariant under the $\, t \, \, \leftrightarrow \, \, 1-t \, \, $
symmetry :
\begin{eqnarray}
\label{Mirror}
(t, \, \sigma, \,  \sigma', \,  \sigma'') \quad \rightarrow \quad \quad 
\Bigl( 1-t,  \, \, \, \, -\sigma \, -1/4, \, \,\, \,   \sigma', 
\,  \,\, \,  -\sigma'' \Bigr)
\end{eqnarray}
This remarkable symmetry is, in fact, inherited from the 
covariance  by (\ref{Mirror}) of the sigma
 form (\ref{jimbo-miwa}) when $\, N\, =0$.

A large set of algebraic solutions of Painlev\'e VI
 (and associated modular curves) 
have been obtained by many authors~\cite{boalch1,boalch2,hit,pic,mazz,hamed}. 
However, most of these results on algebraic solutions 
are for the canonical form\footnote[4]{For $N=0$ this
 equation has been solved in terms
 of theta functions~\cite{hit,pic,mazz}, has dihedral symmetry
 and has a countable number of algebraic solutions.}
 of Painlev\'e VI in
terms of the variable $\, y$:
\begin{eqnarray}
\label{ypan}
&&{d^2y\over dt^2}\, =\,  
{1 \over 2}\left({1\over y}\, +{1\over 1-y}\, +{{1} \over
  {y-t}}\right) \left({dy\over dt}\right)^2-\left({1\over t}\, 
  +{1 \over t-1}\, +{1\over y-t}\right){dy\over dt} \nonumber\\
&&\quad +{y \, (y-1)\, (y-t)\over t^2(t-1)^2}\left(\alpha+\beta{t\over
  y^2}\, +\gamma\, {t-1\over (y-1)^2}\, 
+\delta\, {t \, (t-1)\over (y-t)^2}\right)
\end{eqnarray}
There are several sets of 
$\, \alpha,\, \beta,\, \gamma,\, \delta\, $ which  lead to 
the same equation~\cite{okamoto} for $\, \sigma$. For $\, T < T_c$ one
 such set\footnote[8]{To be considered when comparing with~\cite{hit}.} 
of parameters of (\ref{ypan}), corresponding to the $\, N$-dependent
sigma form (\ref{jimbo-miwa}), is :
\begin{eqnarray}
\label{lesparam2}
\alpha=\, {1 \over 2} \left( N+{1 \over 2} \right)^2,\,\, \,\,  \,
\beta=\, -{1 \over 2} \left(N-{1 \over 2} \right)^2, \,\,\, \,
\gamma=\,{1 \over 8}, \,\,\, \, \,\delta= \,{3 \over 8}\quad\quad \quad\quad
\end{eqnarray}

It is interesting to make the connection between our results and those
previously known algebraic solutions. 
Such a ``dictionary'' will be performed
 elsewhere\footnote[5]{We found that the
$\, SL(2,\, Z)$ subgroup for $\, \tau^n$ for $\, n$ odd,
 (resp. $\, \tau^{2\, n}$ for $\, n$ even)
identify with the one for $\, y$ (see~\cite{Barth,hit}.}, let us just give
 one simple example.  The variable $\, y$ being a 
rational expression of $\, \sigma$ and
its derivatives (see \cite{okamoto}),
the algebraic solution (\ref{c1}) with (\ref{lesparam2}), becomes
 $y=1-\sqrt{1-t}$
which is the well-known solution\footnote[3]{
The solution $\, y=\, \sqrt{t}$ solves (\ref{ypan}) for the parameters
$(\alpha,\, \beta,\, \gamma,\, \delta)\, = \, $
$(\alpha,\, -\alpha,\, 1/2- \delta, \,\delta)$.}
$\, y=\, \sqrt{t}$ (see \cite{hit})
under the change $\, t \rightarrow 1-t$, $y \rightarrow 1-y$,
$\beta \rightarrow -\gamma$ and $\gamma \rightarrow -\beta$ which is a
symmetry of (\ref{ypan}).

\section{Conclusion}
\label{concl}

The diagonal Ising two-point correlation functions
can be expressed (see for instance~\cite{PainleveFuchs,Ghosh2})
 as homogeneous polynomials  
of complete elliptic integral $\, E$ and  $\, K$.
These diagonal Ising correlations are $\lambda=1$ subcase of their
 $\lambda$-extensions $\, C(N,N; \, \lambda)$
we considered in this paper. By (\ref{formm}) and (\ref{formp})
these  polynomials of $\, E$ and $\, K$
are {\em also} expressed as  {\em infinite sums} of the form factors
$\, f^{(j)}_{M,N}$'s which, themselves, are
polynomials of $\, E$ and $\, K$.
 This yields a double infinity $(M, \, N)$ of remarkable identities on the 
complete elliptic integrals $\, E$ and $\, K$. Similarly, with 
the previous algebraic
 solutions for $\, \lambda\, = \, \cos(\pi m/n)$,
one sees that an algebraic expression $\, C(N, N;\cos(\pi m/n))$
 (associated with a modular curve)
can be written as an {\em infinite sum} of polynomials in $\, E$ and $\, K$.
Each of these modular curves will provide a remarkable identity on the 
complete elliptic integrals $\, E$ and $\, K$.

Recalling relations like (5.7)-(5.11) of~\cite{or-ni-gu-pe-01b},
all these  identities can also be written in terms of the 
{\em nome of the elliptic functions} occurring in the Ising model. 
These  identities, now, become remarkable identities on some infinite 
Gaussian sums, or on series expansions
of theta functions, or, for large enough values
 of $ \, N$, on {\em Eisenstein series}
and other {\em quasi-modular forms}. 
We will describe, and analyze these identities in a forthcoming publication.

 The calculations displayed in this paper can be seen as 
successful {\em explicit} examples of {\em factorization
 of multiple integrals}, 
providing examples of explicit calculations
 of the new mantra that  ``nested sums 
are Hopf algebras and thus  multiple Feynman-like integrals must factorize in 
terms of polynomial expressions of 
one-dimensional integrals''. For our $\, j$-particle
contributions of the diagonal correlation 
functions, the $\, C^{(j)}(M, \, N)$'s, the 
fact that they are  polynomial expressions of 
singled-out one-dimensional integrals
(the complete elliptic integrals $\, E$ and $\, K$)
 is understood in terms of direct sums of
linear differential operators equivalent to symmetric powers of a singled-out 
linear differential operator. In the scaling limit, 
this direct sum structure, yielding 
polynomial expressions (that is the so-called  ``factorization 
of multiple integrals''), is lost: what remains
is a Russian-doll structure of differential
 operators equivalent to symmetric powers of a singled-out 
differential operator. 

The problem of the factorization of multiple
 integrals is, obviously, an important one 
for  Feynman-like integrals. It also occurs 
on various calculations of correlation functions of integrable
models, like the Heisenberg spin chain, where 
multiple integrals also occur. 
These factorizations are obtained by Boos and 
Korepin ~\cite{Korepin1,Korepin2} by 
adding to the integrand a successive set of
 anti-symmetric integrands (these anti-symmetric integrands
being chosen in such a way that their multiple 
integral is zero). The combination of 
the initial integral with these new integrands
 yielding expressions depending on less
variables, thus reducing the $n$-multiple
 integrals to a $(n-1)$-multiple integral.  
More recently,  Boos {\it et al.}~\cite{Boos2006} also deduced
factorization of multiple integrals 
representing the density matrix  of the Heisenberg spin chain:
the key ingredient, in the emergence of
 such factorization, is a functional identity
on the integrand, this relation coming from 
the Bethe ansatz integrability of the model.
The factorization of some multiple integrals 
can probably be seen as a consequence of some
``Yang-Baxter integrability'', it seems, 
however, to occur beyond this narrow framework. 
The Feynman-like integrals, where
 such factorization of some multiple integrals
occurs,  are not arbitrary holonomic expressions.
What are the (more or less integrable) 
constraints one must impose on  holonomic integrands 
such that their multiple integrals exhibit factorization, remains
 a fascinating open question~\cite{Kreimer2000}. 
A key point we have tried to promote here is that,
 instead of trying to calculate multiple integrals
where the integrands have no free parameters, that is to say that the
multiple integrals are just constants~\cite{Crandall}, 
we perform calculations on 
multiple integrals where the integrands {\em do depend} on one, or
many, parameters. We can then use the holonomic structure. 

In short: it is simpler to get multiple integrals that depend on one variable
than obtaining their evaluation at a given value on this variable.
This is typically a Yang-Baxter view point : it is
 easier to solve an integrable  model
with a spectral parameter that enables to describe the
Yang-Baxter structure, than trying to solve that
 model for a {\em given value} of that parameter
(quantum groups, knot theory, etc.). It is easier to 
solve the anisotropic Ising model  
than the isotropic one, and, similarly, it is easier to
 consider  {\em multiple integrals that depends 
on a variable, than evaluating constants}~\cite{Kreimer2000} (polynomial 
expressions of $\, \zeta(3))$, $\, \zeta(5)$, ...)
corresponding to these multiple integrals at a given value of that
parameter: this way of looking at the problem 
enables to see the emergence of highly non trivial
 algebraic structures on linear
differential operators, that are a very efficient and powerful tool
of experimental mathematics, and other formal calculations, to study 
factorizations of multiple integrals.

\vskip .5cm 

\textbf{Acknowledgments:}
 We thank Prof. R. J. Baxter for interesting comments 
on this work, and Prof. C.A. Tracy
 for generously providing some of his notes on 
calculations he performed on integrals related to
  the Ising model. We thank A. Its for pointing out
 some important references on
 Picard solutions to PVI and algebraic solutions derived from them. We do 
thank J-A. Weil for illuminating comments on the scaling limits
 of differential operators,  M. Rybowicz
for help in some of our extensive formal calculations, V.E. Korepin 
for  sharing his knowledge of multiple integrable calculations, 
and N. Witte for interesting 
information on the Garnier systems. One of us (BM) has been partially 
supported by NSF grant DMR-0302758.  One of us (BM) thanks 
J-M. Maillet for hospitality at Ecole Normale Sup\'erieure in 
Lyon where part of this work was done.
 One of us (NZ) would like to acknowledge  kind hospitality
at the LPTMC where part of this work has been completed.  
One of us (JMM) thanks Stony Brook and the MASCOS (Melbourne)
 where part of this work was performed.

\vskip .5cm 
\vskip .3cm 

\appendix

\section{Differential operators $L_j(N)$}
\label{B}

The linear differential operators $L_{j}(N)$ have the following form:
\begin{eqnarray}
L_j(N) \, = \, \, \, \sum_{n=0}^{n_0} \, {\frac{N^{2n}}{t^{2n}}}
 \cdot \Bigl( \sum_{k=0}^{j-2k}\,
{\frac{P^{(j)}_{n,k}(t)}{\left(t(t-1) \right)^k}}\, Dt^{j-2n-k} \Bigr)
\end{eqnarray}
where $n_0\, =(j-1)/2$ for $j$ odd, and $n_0=j/2$ for $j$ even.
The polynomials $\, P^{(j)}_{n,k}(t)$ are of degree $\, k$ in $t$.

\subsection{$P^{(5)}_{n,k}(t)$ }
\begin{eqnarray}
&&P^{(5)}_{0,0} =\, 1,\,\, \quad P^{(5)}_{0,1}=\,40\,t-20,\, \quad 
\,P^{(5)}_{0,2}=\, -563\,t+558\,{t}^{2}+118, \nonumber \\
&&P^{(5)}_{0,3} =\, {\frac {4291}{2}}\,t\, 
-{\frac {10169}{2}}\,{t}^{2}+3320\,{t}^{3}-220, \nonumber \\
&&P^{(5)}_{0,4} =\, 80 +10848\,{t}^{2}-16978\,{t}^{3}
+8180\,{t}^{4}-2227\,t, \nonumber \\
&&P^{(5)}_{0,5} =\, 4\, \left(85 -1139\,t+3672\,{t}^{2}
 -4250\,{t}^{3}+ 1600\,{t}^{4}\right)\,t , \nonumber \\
&&P^{(5)}_{1,0} =\, -5,\, \,  \,  P^{(5)}_{1,1} =\, -91\,t+59, \, \,  \, 
P^{(5)}_{1,2}\, =\, -469\,{t}^{2}+626\,t-181, \nonumber \\
&&P^{(5)}_{1,3} =\,  144-840\,t+1368\,{t}^{2}-656\,{t}^{3},\, \
P^{(5)}_{2,0}=\,4,\, \
P^{(5)}_{2,1}=\, 16\,t-16  \nonumber
\end{eqnarray}

\subsection{$P^{(6)}_{n,k}(t)$ }
\begin{eqnarray}
&&  P^{(6)}_{0,0}  = \, 1, \, \,\quad  P^{(6)}_{0,1}=\, 
70\,t-35,\quad P^{(6)}_{0,2}\, =\, 
{\frac {7427}{4}}\,{t}^{2}-{\frac {7427}{4}}\,t+413,  \nonumber \\
&&P^{(6)}_{0,3} =\,  2\, \left( 2\,t-1 \right) 
 \left( 5912\,{t}^{2}-5912\,t+979 \right),\nonumber \\
&&P^{(6)}_{0,4} =\,  {\frac{2410523}{16}}\,{t}^{4}\, 
-{\frac {2410523}{8}}\,{t}^{3}+{\frac {3200163}
{16}}\,{t}^{2}-{\frac {98705}{2}}\,t+3383, \nonumber \\
&&P^{(6)}_{0,5} =\, \,  {1 \over 16}\, \left( 2\,t-1 \right) 
 \left( 3585925\,{t}^{4}-7171850\,{t}^{3}+4326453\,{t}^{2}
-740528\,t+19600 \right), \nonumber \\
&&P^{(6)}_{0,6} =\,  {\frac {625}{64}}\,t \left( t-1 \right) 
 \left( 48841\,{t}^{4}-97682\,{t}^{3}+63549\,{t}^{2}-14708\,t
+784 \right),  \nonumber \\
&&P^{(6)}_{1,0} =\,  -{\frac {35}{4}},\quad P^{(6)}_{1,1}
 = -336\,t+{\frac {413}{2}}, \nonumber \\
&&P^{(6)}_{1,2} =\,  -{\frac {34799}{8}}\,{t}^{2}+{\frac {43231}{8}}\,t
-{\frac {6133}{4}}, \nonumber \\
&&P^{(6)}_{1,3} =\, -{\frac {88609}{4}}\,{t}^{3}+41823\,{t}^{2}
-{\frac {96849}{4}}\,t+{\frac {16691}{4}}, \nonumber \\
&&P^{(6)}_{1,4} =\,  -{\frac {25}{64}}\, \left( t-1 \right)
  \left( 94091\,{t}^{3}-146523\,{t}^{2}+67548
\,t-9216 \right), \nonumber \\
&&P^{(6)}_{2,0} =\,  {\frac {259}{16}},\,\,\quad  P^{(6)}_{2,1}=\,
{\frac {1917}{8}}\,t-{\frac {3159}{16}}, \nonumber \\
&&P^{(6)}_{2,2} =\,  {\frac {125}{64}}\, 
\left( 407\,t-272 \right)  \left( t-1 \right) ,\,
\,\quad P^{(6)}_{3,0} =\, -{\frac {225}{64}} \nonumber 
\end{eqnarray}

\subsection{$P^{(7)}_{n,k}(t)$ }
\begin{eqnarray}
&& P^{(7)}_{0,0}  = 1,\,\,\quad P^{(7)}_{0,1} =-56+112\,t, \,
\quad \,P^{(7)}_{0,2} =\,5012\,{t}^{2}-5026\,t+
1148,  \nonumber \\
&& P^{(7)}_{0,3}  =\, -10736+79727\,t
-174373\,{t}^{2}+115544\,{t}^{3}, \nonumber \\
&& P^{(7)}_{0,4}  =\, 46172-548736\,t+2042953\,{t}^{2}
-2975244\,{t}^{3}+1472828\,{t}^{4}, \nonumber \\
&& P^{(7)}_{0,5}  = \,-78640+1605642\,t-9634279\,{t}^{2}+23975501\,{t}^{3}\, 
 \nonumber \\
&&\quad -26144958\,{t}^{4}  +10305440\,{t}^{5}, \nonumber \\
&&  P^{(7)}_{0,6}  = 29160-1616078\,t
+{\frac {67624527}{4}}\,{t}^{2} \nonumber \\
&&\quad -{\frac {136608085}{2}}\,{t}^{3} +{\frac {511207495}{4}}\,{t}^{4}\, 
-111249042\,{t}^{5}+36334360\,{t}^{6}, \nonumber \\
&& P^{(7)}_{0,7}  = {9 \over 2}\,t \Bigl( 59940-1665037\,t
+11865715\,{t}^{2}-36308026\,{t}^{3} +54466294\,{t}^{4} \nonumber \\
&& \quad -39393900\,{t}^{5}+10951200\,{t}^{6} \Bigr), \quad \nonumber \\
&& P^{(7)}_{1,0}  =\, -14,\quad P^{(7)}_{1,1} =-966\,t+574, \nonumber \\
&& P^{(7)}_{1,2} =\,-24712\,{t}^{2}+29686\,t-8248,\nonumber  \\
&& P^{(7)}_{1,3}  = \,-290812\,{t}^{3}
+530547\,{t}^{2} -299013\,t+51188, \nonumber \\
&& P^{(7)}_{1,4}  = \,-1561136\,{t}^{4}+3851903\,{t}^{3}
-3309480\,{t}^{2} +1156221\,t-136440, \nonumber \\
&& P^{(7)}_{1,5}  =\, 129600-{\frac {22166415}{2}}\,{t}^{3}
+{\frac {18989235}{2}}\,{t}^{4}  
 +{\frac {11893977}{2}}\,{t}^{2}\nonumber \\
&& \quad \quad -{\frac {2902725}{2}}\,t-3028104\,{t}^{5}, \nonumber \\
&&  P^{(7)}_{2,0}  =\, 49,\,\,\quad
 P^{(7)}_{2,1} =\,1686\,t-1254, \quad  P^{(7)}_{2,2} 
=\,17887\,{t}^{2}-27026\,t+9679, \,\,\nonumber \\
&& P^{(7)}_{2,3}  =\, -22761-133569\,{t}^{2}+57753\,{t}^{3} 
 +98253\,t,\,\,\nonumber \\
&& P^{(7)}_{3,0} =\, -36, \qquad P^{(7)}_{3,1} =\, 324-324\,t \nonumber 
\end{eqnarray}

\subsection{$P^{(8)}_{n,k}(t)$ }
\begin{eqnarray}
&& P^{(8)}_{0,0} =\, 1, \quad P^{(8)}_{0,1}=168\,t-84, \,\,\quad
P^{(8)}_{0,2}=\,11697\,{t}^{2}-11697\,t+2730, \nonumber \\
&& P^{(8)}_{0,3} =\, 2\, \left( 2\,t-1 \right)  \left( 109862\,{t}^{2}-
109862\,t+21881 \right), \nonumber \\
&& P^{(8)}_{0,4} = \,{\frac {77675835}{8}}\,{t}^{4}\, 
-{\frac {77675835}{4}}\,{t}^{3}+{\frac {108450015}{8}}\,{t}^{2}\, 
-{\frac {7693545}{2}}\,t+364365, \nonumber \\
&& P^{(8)}_{0,5}  = \,{1 \over 4}\, \left( 2\,t-1 \right)  \Bigl( 
257365313\,{t}^{4}-514730626\,{t}^{3}+340542345\,{t}^{2} \nonumber \\
&&\quad  -83177032\,t+6033464 \Bigr),\,  \nonumber \\
&& P^{(8)}_{0,6}   = \,2610671-{\frac {135579123}{2}}\,t \,
   +{\frac {4351723053}{8}}\,{t}^{2}
-{\frac {31150612733}{16}}\,{t}^{3} \nonumber \\
&&\quad -{\frac {47738959467}{16}}\,{t}^{5}+{\frac {55357772589}{16}}\,{t}^{4} 
 +{\frac {15912986489}{16}}\,{t}^{6},  \nonumber \\
&& P^{(8)}_{0,7}   =\, {1 \over 8}\, \left( 2\,t-1 \right) 
 \Bigl( 16309728941\,{t}^{6}-48929186823\,{t}^{5} 
 +54824769942\,{t}^{4} \nonumber \\
 &&\quad -28100895179\,{t}^{3}+6440184015\,{t}^{2}-544600896\,t 
  +8016008 \Bigr),    \nonumber \\
&& P^{(8)}_{0,8}   =\, {\frac {2401}{256}}\,t \left( t-1 \right) 
 \Bigl( 719580625\,{t}^{6}-2158741875\,{t}^{5}
 +2496751275\,{t}^{4}  \nonumber \\
&&\quad    -1395599425\,{t}^{3}+383051976\,{t}^{2}
-45042576\,t+1308736 \Bigr), \nonumber \\
&& P^{(8)}_{1,0}   =\, -21, \, \, \, \quad P^{(8)}_{1,1} \, 
 =\, \, -2352\,t+1365, \quad \, \, \, \nonumber \\
&& P^{(8)}_{1,2}   = \,-{\frac {414555}{4}}\,{t}^{2}
+{\frac {483843}{4}}\,t  -33315, \,\,\nonumber \\
&& P^{(8)}_{1,3}   =\, -2290461\,{t}^{3}+4034358\,{t}^{2}
-2237787\,t+386664, \nonumber \\
&& P^{(8)}_{1,4}   =\, -{\frac {426526863}{16}}
\,{t}^{4}+{\frac {504203159}{8}}\,{t}^{3}\, 
-{\frac {845513895}{16}}\,{t}^{2}+{\frac {36865265}{2}}\,t \nonumber \\
&& \quad -2230431,  \nonumber \\
&& P^{(8)}_{1,5}  = -{\frac {616586181}{4
}}\,{t}^{5}+{\frac {7342474719}{16}}\,{t}^{4}-{\frac {4139827129}{8}}
\,{t}^{3} \nonumber \\
&& \quad +{\frac {4378085671}{16}}\,{t}^{2}-67155042\,t+6072033, \nonumber \\
&& P^{(8)}_{1,6}  = \,-{\frac {49}{64}}\, 
\left( t-1 \right)  \Bigl( 449304249\,{t}^{5}
-1168884874\,{t}^{4}+1134316077\,{t}^{3} \nonumber \\
&&\quad  -509448428\,{t}^{2}+105774112\,t-8294400 \Bigr), \nonumber \\
&&  P^{(8)}_{2,0} =\, {\frac {987}{8}},  \qquad
 P^{(8)}_{2,1}=\,{\frac {15993}{2}}\,t-{\frac {22299}{4}}, \nonumber \\
&& P^{(8)}_{2,2}=\,{\frac {2933043}{16}}\,{t}^{2}\,
-{\frac {4128099}{16}}\,t+{\frac {696405}{8}}, \nonumber \\
&& P^{(8)}_{2,3} = \,{\frac {
7002915}{4}}\,{t}^{3}-{\frac {14949545}{4}}\,{t}^{2}+{\frac {10236397}
{4}}\,t-{\frac {4465707}{8}}, \nonumber \\
&& P^{(8)}_{2,4} = \,{\frac {343}{128}}\, \left( t-1 \right) 
 \left( 2179797\,{t}^{3}-4103797\,{t}^{2}+2457908\,t-468864
 \right), \nonumber \\
&& P^{(8)}_{3,0}  =\, -{\frac {3229}{16}}, \,\, \quad  P^{(8)}_{3,1}=\, 
-{\frac {21963}{4}}\,t+{\frac {76827}{16}}, \nonumber \\
&& P^{(8)}_{3,2} =\, -{\frac {343}{64}}\, \left( 6607\,
t-5032 \right)  \left( t-1 \right),\, \, \,  \, \, \quad 
P^{(8)}_{4,0}={\frac {11025}{256}}  \nonumber
\end{eqnarray}

\subsection{$P^{(9)}_{n,k}(t)$ }
\begin{eqnarray}
&& P^{(9)}_{0,0}  = \,1, \,\, \quad P^{(9)}_{0,1} =\, -120+240\,t,\,\,\quad
  P^{(9)}_{0,2} =\, 5796-24546\,t+24516\,{t}^{2}, \nonumber \\
&& P^{(9)}_{0,3}  = \,-145528+991701\,t
-2099751\,{t}^{2}+1396208\,{t}^{3}, \nonumber \\
&& P^{(9)}_{0,4} =\, 2045004-20325858\,t+69369177\,{t}^{2}
-97902648\,{t}^{3}+48749364\,{t}^{4}, \nonumber \\
&& P^{(9)}_{0,5} =\, -16074560+225525578\,t-1125696965\,{t}^{2
}+2565535675\,{t}^{3} \nonumber \\
&&\quad  -2714936962\,{t}^{4}+1079617840\,{t}^{5}, \nonumber \\
&& P^{(9)}_{0,6} =\, 66126712-1333788966\,t
+{\frac {37765468163}{4}}\,{t}^{2}-{\frac 
{63124281313}{2}}\,{t}^{3} \nonumber \\
&&\quad   +{\frac {216166206483}{4}}\,{t}^{4}
-45734526046\,{t}^{5}+15125870712\,{t}^{6}, \nonumber \\
&& P^{(9)}_{0,7} =\, -118102672+3823928460\,t-{\frac {78510959875}{2}}\,{t}^{2}
+{\frac {374049548401}{2}}\,{t}^{3} \nonumber \\
&&\quad  -471178501099\,{t}^{4}+646530989251\,{t}^{5}-455734056216
\,{t}^{6}  \nonumber \\
&&\quad  +128906004992\,{t}^{7}, 
 P^{(9)}_{0,8} = \,47071232-4139526516\,t+{
\frac {138902716891}{2}}\,{t}^{2} \nonumber \\
&&\quad   -484196478836\,{t}^{3} 
 +{\frac {3495148889889}{2}}\,{t}^{4}-3539969007392\,{t}^{5} \nonumber \\
&&\quad   +4054878125399\,{t}^{6} 
 -2448333931344\,{t}^{7}+604418968592\,{t}^{8}, \nonumber \\
&& P^{(9)}_{0,9} =\, 80\,t \Bigl( 9561344-427020633\,t
+4937178194\,{t}^{2}-26308505171\,{t}^{3} \nonumber \\
&&\quad   +76760779797\,{t}^{4}-130255661861\,{t}^{5}
 +128108854250\,{t}^{6} \nonumber \\
&&\quad   -67626000000\,{t}^{7}+14796800000\,{t}^{8} \Bigr), \,\,\quad
 P^{(9)}_{1,0}  =\, -30, \nonumber \\
&& P^{(9)}_{1,1} =\,-5082\,t+2898, \, \, P^{(9)}_{1,2} =\, 
-352662\,{t}^{2}+404466\,t-110238, \nonumber \\
&& P^{(9)}_{1,3} = \,-12963996\,{t}^{3}+22438245\,{t}^{2}
-12306435\,t+2123604, \nonumber \\
&& P^{(9)}_{1,4} =\, -271930980\,{t}^{4}+631696597\,{t}^{3}
-523169724\,{t}^{2}+181823257\,t  \nonumber \\
&& \quad -22193940, \, \,
 P^{(9)}_{1,5} =\,  -3245449704\,{t}^{5}
+{\frac {18983501249}{2}}\,{t}^{4}  \nonumber \\
&& \quad -{\frac {21116262613}{2}}\,{t}^{3} 
 +{\frac {11093266991}{2}}\,{t}^{2}-{\frac {2731643299}{2}}\,t
+125146416, \nonumber \\
&& P^{(9)}_{1,6} = \, -20342103432\,{t}^{6}
+71951600804\,{t}^{5}-100847772344\,{t}^{4} \nonumber \\
&&\quad +71245445309\,{t}^{3}-26573340926\,{t}^{2}+4930067225\,t
-354631488, \nonumber \\
&& P^{(9)}_{1,7} =\,  406425600-363525018400\,{t}^{5}
-6762200560\,t+46013156464\,{t}^{2} \nonumber \\
&& \quad+214239244800\,{t}^{6}-51475353600\,{t}^{7}
+324098542224\,{t}^{4}  \nonumber \\
&& \quad-162977694704\,{t}^{3}, \quad
 P^{(9)}_{2,0}  =\,  273, \quad
P^{(9)}_{2,1} =\,  29490\,t-19650, \nonumber \\
&& P^{(9)}_{2,2} =\, 1217265\,{t}^{2}-1636902\,t+528465, \nonumber \\
&& P^{(9)}_{2,3} =\,  23917695\,{t}^{3}-
48731759\,{t}^{2}+31834675\,t-6637935, \nonumber \\
&& P^{(9)}_{2,4} = \, 222934641\,{t}^{4}-612371540\,{t}^{3}
+607781638\,{t}^{2}-257369288\,t \nonumber \\
&&\quad +39119361, 
 P^{(9)}_{2,5} =\,  734599360\,t-2367055040\,{t}^{2}
-2730725376\,{t}^{4} \nonumber \\
&&\quad -87745536 
 +785703936\,{t}^{5}+3664705792\,{t}^{3}, \nonumber \\
&& P^{(9)}_{3,0}  = \, -820, \, \, 
P^{(9)}_{3,1} =\, -46428\,t+37212, \nonumber \\
&& P^{(9)}_{3,2} = \, -839284\,{t}^{2}+1358312\,t-535156, \nonumber \\
&& P^{(9)}_{3,3} = \, 2455552-9463296\,t+11831808\,{t}^{2}
-4814848\,{t}^{3}, \nonumber \\
&& P^{(9)}_{4,0} =\,  576, \, \, \, \,  P^{(9)}_{4,1} =\,
 -9216+9216\,t  \nonumber
\end{eqnarray}

\subsection{$P^{(10)}_{n,k}(t)$ }
\begin{eqnarray}
&& P^{(10)}_{0,0} =\, \,  1,\, \, P^{(10)}_{0,1} =-165+330\,t, \, \, 
P^{(10)}_{0,2} =\, \, 11286-{\frac {189189}{4}}\,t
+{\frac {189189}{4}}\,{t}^{2}, \nonumber \\
&& P^{(10)}_{0,3} = \, \, 440\, \left( -1+2\,t \right) 
 \left( 4400\,{t}^{2}-4400\,t+947 \right), \nonumber \\
&& P^{(10)}_{0,4} = 9053979-85922628\,t
+{\frac {2289151821}{8}}\,{t}^{2}-{\frac {1601770797}{4}}
\,{t}^{3}  \nonumber \\
&&\quad  +{\frac {1601770797}{8}}\,{t}^{4}, \, \,
\nonumber \\
&& P^{(10)}_{0,5} =\, \,  {1 \over 8}\, \left( 2\,t-1 \right) 
 \Bigl( 27291921049\,{t}^{4}-54583842098\,{t}^{3} \nonumber \\
 &&\quad  +37651935321\,{t}^{2}-10360014272\,t+946138408 \Bigr), \nonumber \\
&& P^{(10)}_{0,6} = \, \, 907059937-{\frac {63513668189}{4}}\,t
+{\frac {414126483423}{4}}\,{t}^{2}  \nonumber \\
&&\quad  -{\frac {10567258749853}{32}}\,{t}^{3}
+{\frac {17677263640199}{32}}\,{t}^{4}
-{\frac {14872361118327}{32}}\,{t}^{5}  \nonumber \\
&&\quad  +{\frac {4957453706109}{32}}\,{t}^{6}, \, \, 
\nonumber \\
&& P^{(10)}_{0,7} =\, {1 \over 8}\, \left(2\,t-1 \right)
  \Bigl( 9240801571631\,{t}^{6}  \nonumber \\
&& \quad -27722404714893\,{t}^{5}+32221157315067\,{t}^{4}
-18238306771979\,{t}^{3} \nonumber \\
&&  \quad +5175369000414\,{t}^{2}-676616400240\,t+
30201789392 \Bigr), \nonumber \\
&& P^{(10)}_{0,8}  = 7010881775-{\frac {537428072635}{2}}\,
t+{\frac {53588823341945}{16}}\,{t}^{2} \nonumber \\
&&\quad -{\frac {320491542697265}{16}}\,{t}^{3}
+{\frac {16973016403001045}{256}}\,{t}^{4} \nonumber \\
&&\quad -{\frac {8195527196507945}{64}}\,{t}^{5} \, 
 +{\frac {18373024724608855}{128}}\,{t}^{6}  \nonumber \\
&&\quad -{\frac {5532574254401525}{64}}\,{t}^{7}
+{\frac {5532574254401525}{256}}\,{t}^{8}, \nonumber \\
&& P^{(10)}_{0,9} = \, \, {\frac {5}{256}}\, \left( 2\,t-1 \right)
 \Bigl( 2925753951778285\,{t}^{8}-11703015807113140\,{t}^{7}  \nonumber \\
&&\quad +19103849088522126\,{t}^{6}-16350991940670388\,{t}^{5} \nonumber \\
&&\quad +7838575034697949\,{t}^{4}-2079015276577248\,{t}^{3}
+280113055050736\,{t}^{2} \nonumber \\
&&\quad -15268105688320\,t+148553637120 \Bigr), \nonumber \\
&& P^{(10)}_{0,10} =\, \,  {\frac {32805}{1024}}\,t \left( t-1 \right)
  \Bigl( 8079810760125\,{t}^{8}-32319243040500\,{t}^{7} \nonumber \\
&&\quad +53723369995078\,{t}^{6}-48052759343484\,{t}^{5}
+24975072368117\,{t}^{4} \nonumber \\
&&\quad  -7567996044344\,{t}^{3}+1257903576048\,{t}^{2}
-96158271040\,t \nonumber \\
&&\quad  +1833995520 \Bigr),\, \, \quad 
P^{(10)}_{1,0} = \, -{\frac {165}{4}},  \,  \, \quad 
P^{(10)}_{1,1} = \, -10032\,t+5643, \nonumber \\
&& P^{(10)}_{1,2} = \, -{\frac {4116057}{4}}\,{t}^{2}+{
\frac {4646169}{4}}\,t-{\frac {629409}{2}}, \nonumber \\
&& P^{(10)}_{1,3} = \, -{\frac {116107101}{2}}\,{t}^{3}+98646966\,{t}^{2}\, 
-{\frac {107291349}{2}}\,t+{\frac {18577449}{2}},  \nonumber \\
&& P^{(10)}_{1,4} = \, -{\frac {62944154655}{32}}\,{t}^{4}+{
\frac {71575322887}{16}}\,{t}^{3}
-{\frac {117218126643}{32}}\,{t}^{2} \nonumber \\
&&\quad  +{\frac {10187763521}{8}}\,t
-{\frac {630658425}{4}}, \,  \, \nonumber \\
&&P^{(10)}_{1,5} = \, -{\frac{163658617341}{4}}\,{t}^{5} 
 +{\frac {1868493139019}{16}}\,{t}^{4}
-{\frac {1024185567025}{8}}\,{t}^{3} \nonumber \\
&& \quad  +{\frac {1072700805259}{16}}\,{t}^{2}  
 -16676053189\,t+{\frac {3132363327}{2}}, \nonumber \\
&& P^{(10)}_{1,6} = \, -{\frac {32591037777225}{64}}\,{t}^{6} 
 +{\frac {112104944104795}{64}}\,{t}^{5}-{\frac {154297814907493}{64}}\,{t}^{4}
\nonumber \\
&&\quad +{\frac {108234764940653}{64}}\,{t}^{3}  
-{\frac {20292699152369}{32}}\,{t}^{2}\nonumber \\
&&\quad +{\frac {958545911705}{8}}\,t
-{\frac {35460987675}{4}}, \nonumber \\
&& P^{(10)}_{1,7} =-{\frac {110727690476325}{32}}\,{t}^{7}
+{\frac {446374095368415}{32}}\,{t}^{6} \nonumber \\
&&\quad -{\frac {740746002842197}{32}}\,{t}^{5}
 +{\frac {163197508324913}{8}}\,{t}^{4}-{
\frac {82052435490193}{8}}\,{t}^{3}  \nonumber \\
&&\quad +{\frac {93651745559635}{32}}\,{t}^{2} 
 -{\frac {1748299590545}{4}}\,t+{\frac {105491089125}{4}},  \nonumber \\
&& P^{(10)}_{1,8}  = -{\frac {6561}{1024}}\, \left( t-1 \right) 
 \Bigl( 1530421397125\,{t}^{7}   -5554611547375\,{t}^{6}  \nonumber \\
&&\quad +8232623167111\,{t}^{5}-6428537243541\,{t}^{4}
+2844044623496\,{t}^{3} \nonumber \\
&&\quad -711552088080\,{t}^{2}+93740238400\,t-5138022400 \Bigr), \nonumber \\
&& P^{(10)}_{2,0} =\,  {\frac {4389}{8}},\,\, \qquad 
P^{(10)}_{2,1} =\, {\frac {364353}{4}}\,t-{\frac {469491}{8}}, \nonumber \\
&& P^{(10)}_{2,2} =\, {\frac {194340135}{32}}
\,{t}^{2}-{\frac {251951799}{32}}\,t+{\frac {19747365}{8}}, \nonumber \\
&& P^{(10)}_{2,3} = \,{\frac {830796045}{4}}\,{t}^{3}
-406585301\,{t}^{2}+{\frac {1027545163}
{4}}\,t\, -{\frac {208821765}{4}}, \nonumber \\
&& P^{(10)}_{2,4} = \,{\frac {490847729943}{128}}
\,{t}^{4}-{\frac {645152215543}{64}}\,{t}^{3}
+{\frac {1232916142207}{128}}\,{t}^{2} \nonumber \\
&& \quad -{\frac {63246182437}{16}}\,t+{\frac {4689902523}{8}},\nonumber \\
&&P^{(10)}_{2,5} =\,{\frac {2314309478331}{64}}\,{t}^{5}
 -{\frac {15325677165945}{128}}\,{t}^{4}
+{\frac {1231151157175}{8}}\,{t}^{3}\nonumber \\
&&\quad -{\frac {12252529589789}{128}}\,{t}^{2}  \,
 +{\frac {230003188957}{8}}\,t-{\frac {26662508757}{8}}, \nonumber \\
&& P^{(10)}_{2,6}\, =\,{\frac {729}{512}}\, \left( t-1 \right)
 \Bigl( 95338644413\,{t}^{5}-286574346250\,{t}^{4}  \nonumber \\
&& \quad +332882516705\,{t}^{3}-186665173556\,{t}^{2}
+50609686768\,t  -5337817088 \Bigr), \nonumber \\
&&P^{(10)}_{3,0} =\, -{\frac {86405}{32}}, \qquad 
P^{(10)}_{3,1} =\, -{\frac {1071807}{4}}\,t
+{\frac {3230007}{16}},  \nonumber \\
&& P^{(10)}_{3,2} =\, -{\frac {616031665}{64}}\,{t}^{2}\,
+{\frac {934953233}{64}}\,t-{\frac {173220155}{32}}, \nonumber \\
&& P^{(10)}_{3,3} =\, -{\frac {4731457901}{32}}\,{t}^{3}
+{\frac {2715664857}{8}}\,{t}^{2}-{\frac {
8133623529}{32}}\,t+{\frac {1981312349}{32}},  \nonumber \\
&& P^{(10)}_{3,4} =\, -{\frac {2187}{512}}\, \left( t-1 \right) 
 \Bigl( 191281007\,{t}^{3}-399820191\,{t}^{2}+271766508\,t  \nonumber \\
&&\quad  -60099968 \Bigr),\,\,\nonumber \\
&&P^{(10)}_{4,0} =\, {\frac {1057221}{256}},\,\,\,\,\quad 
P^{(10)}_{4,1} =\, {\frac {23053617}{128}}\,t
-{\frac {41642109}{256}},  \nonumber \\
&& P^{(10)}_{4,2} =\, {\frac {2187}{1024}}\, \left( 892447\,t-727072 \right) 
 \left( t-1 \right), \,\, \quad P^{(10)}_{5,0} =\,
 -{\frac {893025}{1024}} \nonumber
\end{eqnarray}

\section{Direct sum structure}
\label{C}

We display the fourth order differential operator $\, M_4(N)$
introduced in sec. \ref{beyond} 
for successive values of $\, N$:
\begin{eqnarray}
&&M_4(0) \, = \, \, Dt^{4}\,\,
 +2\,{\frac { \left( 2\,t-1 \right)  \left( 2\,{t}^{2}-2
\,t+3 \right) }{  \left( {t}^{2}-t+1
 \right)\, \left( t-1 \right)\,  t}} \cdot  Dt^{3} \nonumber \\
&&\quad  +{1 \over 2}\,{\frac { \left( -73\,t+14+102\,{t}^{2}-58\,{t}^{3}+
29\,{t}^{4} \right) }{
 \left( {t}^{2}-t+1 \right) \, \left( t -1 \right)^{2} \, t^2}}
 \cdot Dt^2 \nonumber \\
&&\quad  +{1 \over 2}\,{\frac { \left(2\,t-1 \right)
 \left( 5\,{t}^{4}-10\,{t}^{3}+27\,{t}^{2}-22\,t+2 \right)  }{
\left( {t}^{2}-t+1 \right)\, ( t-1)^{3}\,  t^{3}  }} \cdot Dt \nonumber \\
&&\quad  +{1 \over 16}\,{\frac {{t}^{4}-2\,{t}^{3}+42\,{t}^{2}-41\,t+4}{ 
 \left( {t}^{2}-t+1 \right)\, ( t-1)^{3}\,  t^{3}  }} \nonumber
\end{eqnarray}

\begin{eqnarray}
&&M_4(1)\, =\, \,
 Dt^{4}\, +2\,{\frac { P_3 }{ \left( t-1 \right)\, t \cdot  P_4}} \cdot  Dt^3
 \, \,+{1 \over 2}\,{\frac { P_2\,}
{ \left( t-1 \right)^{2} t^2 \, P_4}}\cdot  Dt^2 \nonumber \\
&&\quad \quad \quad  +{\frac { P_1 }
{ \left(t-1 \right)^3\, t^{3} \cdot  P_4}} \cdot  Dt\, \,\,\,
\, +{1 \over 16}\,{\frac { P_0}
{ \left( t-1 \right)^{3} \,{t}^{4} \cdot  P_4}} \nonumber
\end{eqnarray}
where
\begin{eqnarray}
&&P_0 \, =\,  256\,{t}^{6}-560\,{t}^{5}+312\,{t}^{4}
-143\,{t}^{3}+227\,{t}^{2}-72\,t \, -16,    \nonumber  \\
&&P_1 \, =\,   64\,{t}^{7}-856\,{t}^{6}+2826\,{t}^{5}-4087\,{t}^{4}\, 
+2978\,{t}^{3}-1098\,{t}^{2}+182\,t-8,    \nonumber  \\
&&P_2 \, =\,   208+7807\,{t}^{2}-14253\,{t}^{3}+12412\,{t}^{4}
-4624\,{t}^{5}+448\,{t}^{6},    \nonumber  \\
&&P_3 \, =\, \,  64\,{t}^{5}-556\,{t}^{4}
+1225\,{t}^{3} -1078\,{t}^{2}+396\,t-48,    \nonumber  \\
&&P_4 \, =\,  16+209\,{t}^{2}-120\,{t}^{3}+16\,{t}^{4}-120\,t,  \nonumber
\end{eqnarray}

\begin{eqnarray}
&&M_4(2) \, =\, \,  
 Dt^{4}\,\, +2\,{\frac { P_3\, }{t \left( t-1 \right) \,  P_4}} \cdot Dt^3
\, +{1 \over 2}\,{\frac {  P_2\, }{ t^{2}
 \left( t-1 \right)^{2}\, P_4}} \cdot Dt^2 \nonumber \\
&&\quad \quad +{1 \over 2}\,{\frac { P_1\,}{ t^{3}
 \left(t-1 \right)^{3}\, P_4}} \cdot Dt \,\, 
+{1 \over 16}\,{\frac { P_0}{t^4 \left(t-1 \right)^{3} \, P_4}} \nonumber
\end{eqnarray}
where the corresponding $\, P_i$'s read : 
\begin{eqnarray}
&&P_0 \, =\,  
-1344\,{t}^{11}+10752\,{t}^{10}+139321\,{t}^{9}-721147\,{t}^{8}+
1888781\,{t}^{7} \nonumber \\
&& \quad -3452437\,{t}^{6}+4219535\,{t}^{5}
-3184189\,{t}^{4}+1330028\,{t}^{3} \nonumber \\
&&\quad  -202384\,{t}^{2}-34048\,t+7168,    \nonumber  \\
&&P_1 \, =\,   448\,{t}^{11}+4256\,{t}^{10}+56658\,{t}^{9}
-519911\,{t}^{8}+1502563\,t^{7} -2077796\,{t}^{6} \nonumber \\
&&\quad  +1426525\,{t}^{5}-372047\,{t}^{4}-39536\,{t}^{3
}+5418\,{t}^{2}+14336\,t-896,    \nonumber  \\
&&P_2 \, =\, \,   4928\,{t}^{10}-37632\,t-1394407\,{t}^{3}
+4810853\,{t}^{4}-8001289\,{t}^{5} \nonumber \\
&& \quad +6880493\,{t}^{6}+415793\,{t}^{8}-2881207\,{t}^{7}
+16128\,{t}^{9}+11648  \nonumber \\
&&\quad  +174818\,{t}^{2},    \nonumber  \\
&&P_3 \, =\,   1344\,{t}^{9}+1568\,{t}^{8}+65828\,{t}^{7}
-382102\,{t}^{6}+760238\,{t}^{5}-702181\,{t}^{4} \nonumber \\
&&\quad  +302183\,{t}^{3}-46627\,{t}^{2}+1568\,t-1792,    \nonumber  \\
&&P_4 \, =\,  448\,{t}^{8}+448\,{t}^{7}+16513\,{t}^{6}
-81242\,{t}^{5}+127675\,{t}^{4}-81242\,{t}^{3} \nonumber \\
&&\quad  +16513\,{t}^{2}+448\,t+448   \nonumber
\end{eqnarray}

\section{The form factors $\, f^{(j)}_{N,N}$}
\label{AA}

In order to check all the results displayed in this paper, we have performed
a large number of series expansions. Even the series expansions obtained
recursively, order by order, from the sigma form of  Painlev{\'e} VI
(\ref{jimbo-miwa}) in sec. 3, were checked against series expansions
obtained independently.
Some were based on extremely large series expansions, not in $\, s$ or $\, t$,
but in the nome of elliptic 
functions (see (5.7)-(5.11) of~\cite{or-ni-gu-pe-01b}),
others were obtained from series expansions with hypergeometric functions
coefficients.

Actually our new simple integral representations (\ref{2n}),
(\ref{2n-1}) are of a great help to produce large series expansions for 
the quantities  $\, f_{N,N}^{(2n)}(t) $ and $\, f_{N,N}^{(2n+1)}(t)$.
This amounts  to expanding {\em only } 
the  $ \,  (1\, -t\, x_{2j-1} x_{2k})^{-2}$ term 
in (\ref{2n-1}). Recalling the  Euler representation
 of the hypergeometric functions~\cite{Erde}:
\begin{eqnarray}
&&F(a, b, \, c ; t) \,\, = \,\, {{\Gamma(c) } 
\over {\Gamma(c-b) \, \Gamma(b)  }}  \\
&&\qquad \quad  \int_0^{1}  \, x^{b-1} \, 
(1-x)^{c-b-1} \, (1\, -x\, t)^{-a} \, dx \nonumber
\end{eqnarray}
  one can rewrite, alternatively, these integral representations (\ref{2n}),
(\ref{2n-1}) expansions of 
$\, f_{N,N}^{(2n)}(t) $ and $\, f_{N,N}^{(2n+1)}(t)$, as nested sums of 
products of hypergeometric functions. 
By expanding the factor $ \,(1\, -t\, x_1x_2)^{-2} \,$ in a power series
 in $ \,t\, $ we obtain
\begin{eqnarray}
\label{nF2}
&&f_{N,N}^{(2)}(t) \, = \, \,\,  \,
 t^{N+1}\cdot \sum_{j=0}^{\infty} \,(j+1) \, \, t^j \cdot 
{(1/2)_{N+j}(3/2)_{N+j}\over 4
  (N+j+1)!^2}  \, \times      \nonumber  \\
&& \, F(-1/2,N+j+1/2,N+j+2;t) \, F(1/2,N+j+3/2,N+j+2;t)  \qquad \nonumber 
\end{eqnarray}

The series expansions for
$\,  h_{2j}(N, \, N)(t)$'s and the $\,  h_{2j+1}(N, \, N)(t)$'s
agree with the series expansions for the $ \,\hat{C}^{j}(N, \, N)$'s
 and with the series expansions for the
$f^{(2j)}_{N,N}$'s and  $f^{(2j+1)}_{N,N}$'s.
In this Appendix, we display the
$f^{(2j)}_{N,N}$'s and  $f^{(2j+1)}_{N,N}$'s for some $j$ and some $N$.

\subsection{$f^{(1)}_{N,N}$ and $f^{(2)}_{N,N}$}
The $f^{(1)}_{N,N}$'s are given in the text by (\ref{sol1}) and (\ref{hyper}).
The $f^{(2)}_{N,N}$'s are given explicitly as a function of $K$ and $E$
in the text.

\subsection{$f^{(3)}_{N,N}$}
The $\, f^{(3)}_{N,N}$'s read for $\, N=\, 0, \cdots, 4$:
\begin{eqnarray}
&& 6 \,  f^{(3)}_{0,0} \,  = \, \, 
 K  -\left( t-2 \right)\, {K}^{3}\, \, -3\,{K}^{2}\,E \nonumber \\
&& 6\, t^{1/2} \,\,  f^{3}_{1,1} \, =\, \,   4\,(K-\,E) \, 
-6\,{K}^{2}\, E \, - \left( 2\,t-3 \right)\, {K}^{3}\,
+3\,K\,{E}^{2} \nonumber \\
&& 18\, t \,\, f^{(3)}_{2,2} \,\,  = \, \, \, 
7\, \left( t+2 \right)\,  K \, \, 
 -14\, \left( t+1 \right)\,  E\,+ 24\,{E}^{3}\,
  \nonumber \\
&&\quad  \, +3\, \left( 2\,{t}^{2}-11\,t+2 \right)\, E\,{K}^{2}\,\, 
-3\, \left( {t}^{2}-2 \right)\, {K}^{3}\,
 +36\, \left( t-1 \right) K{E}^{2} \nonumber \\
&& 270 \, \, t^{5/2} \, \,  f^{(3)}_{3,3} \, =\, 
 -30\, \left( 8\,{t}^{2}+7\,t+8 \right)\,\,t \cdot   E \,\,
 +30 \left( 4\,{t}^{2}+3\,t+8 \right) K \nonumber \\
&&\quad - \left( 72\,{t}^{4}-158\,{t}^{3}+189\,{t}^{2}
-156\,t+8 \right)\,  {K}^{3} \nonumber \\
&&\quad  +6\, \left( 24\,{t}^{4}-108\,{t}^{3}+29\,{t}^{2}
-6\,t+4 \right) \, E\,{K}^{2} \nonumber \\
&&\quad  +3\, \left( 232\,{t}^{3}-111\,{t}^{2}
-180\,t-8 \right)\,  {E}^{2}\, K \nonumber \\
&&\quad  +4\, \left( t+1 \right)  \left( 2\,{t}^{2}
+103\,t+2 \right)\,t \,  {E}^{3}  \nonumber \\
&& 47250\, t^4\,\,  f^{(3)}_{4,4}\,  \, = \, \, \,   975\,\, 
 \left( 3\,t+4 \right)  \left( 8\,{t}^{2}-5\,t
+12 \right)\,{t}^{2}\,  K \nonumber \\
&&\quad  -7800\,{t}^{2} \left( t+1 \right)
  \left(6\,{t}^{2}-t+6 \right)\,  E  \nonumber \\
&&\quad  - \Bigl( 16216\,{t}^{6}-32109\,{t}^{5}+4218\,{t}^{4}
+38472\,{t}^{3}-38064\,{t}^{2}  +3264\,t+128 \Bigr)\, {K}^{3}\, 
\nonumber \\
&&\quad +3\, \Bigl( 10832\,{t}^{6}-43424\,{t}^{5}+4925\,{t}^{4}
  +13248\,{t}^{3}-10112\,{t}^{2}+3328\,t+128
 \Bigr)\,  E\,{K}^{2} \nonumber \\
&&\quad  -48\, \left( 4\,{t}^{6}-2885\,{t}^{5}+939\,{t}^{4}
+1510\,{t}^{3}+1792\,{t}^{2}+212\,t+8 \right)\, {E}^{2}\, K \nonumber \\
&&\quad  +16\, \left( 8\,{t}^{6}+216\,{t}^{5}+4893\,{t}^{4}
+5464\,{t}^{3}+4893\,{t}^{2}+216\,t+8 \right)\, {E}^{3} \nonumber 
\end{eqnarray}

\subsection{$f^{(4)}_{N,N}$}
Some of the $f^{(4)}_{N,N}$'s read:
\begin{eqnarray}
&& 24\, f^{(4)}_{0,0} \,\,   =\,\,\,  4\, (K-E)\cdot K \, \, 
- \left( 2\,t-3 \right) {K}^{4}-6\,{K}^{3}E\,+3\,{K}^{2}{E}^{2}\, \nonumber \\
&& 24\, f^{(4)}_{1,1} \,\,   =\,\, 
9\, \, -30\,K\, E\, \,  -10\, \left(t-2 \right) \,{K}^{2} \nonumber \\
&&\quad \, + \left( {t}^{2}-6\,t+6 \right) {K}^{4}\,+15\,{K}^{2}{E}^{2}\, 
  +10\, \left(t-2 \right)\, {K}^{3}\,E\,   \nonumber \\
&& 72\, t \cdot  f^{(4)}_{2,2} \,\,   = \, \, \, 
72\,t    -32\, \left( 1+t \right)\, {E}^{2}
\nonumber \\
&& \quad
 -16\, \left( 2+6\,{t}^{2}-11\,t \right)\, {K}^{2}
\, \, -16\, \left(15\,t-4 \right) \, K\, E\nonumber \\
&&\quad 
+ \left( 24\,{t}^{3}-98\,{t}^{2} +113\,t-36 \right)\,  {K}^{4}\, 
\nonumber \\
&&\quad   +12\, \left( 9+t \right)\,  {E}^{3}\, K\, 
+3\, (71\,t -60)\,  {K}^{2}{E}^{2} \nonumber \\
&& \quad \, +2\, \left( 66+74\,{t}^{2}
-157\,t \right) E{K}^{3}\, -24\,{E}^{4}  \nonumber \\
&&1080\, t^2\, f^{(4)}_{3,3} \,\,   =\, \, \,
 22\, \left( 8\,{t}^{3}-319\,{t}^{2}+112\,t+16 \right)\,  KE
\nonumber \\
&&\quad -88\, \left( 1+t \right)  
\left( 2\,{t}^{2}+13\,t+2 \right)\,  {E}^{2}  \nonumber \\
&&\quad  + \left( 957\,{t}^{4}-3646\,{t}^{3}
+4230\,{t}^{2}-1488\,t-8 \right) {K}^{4} \nonumber \\
&&\quad  +8\, \left(46\,{t}^{3}
+51\,{t}^{2}+543\,t -110 \right) {E}^{3}K \nonumber \\
&&\quad  +3\, \left( 16\,{t}^{4}-72\,{t}^{3}
+2537\,{t}^{2}-2704\,t+272 \right) {K}^{2}{E}^{2} \nonumber \\
&&\quad  -6\, \left( 8\,{t}^{4}-903\,{t}^{3}+1934\,{t}^{2}
-988\,t+40 \right) E{K}^{3} \nonumber \\
&&\quad  +24\, \left( 13+13\,{t}^{2}-28\,t \right) {E}^{4} \nonumber \\
&&\quad  -22\, \left( 137\,{t}^{3}-242\,{t}^{2}+52\,t
+8 \right) {K}^{2}+2025\,{t}^{2} \nonumber
\end{eqnarray}

\subsection{$f^{(5)}_{N,N}$}
We give some $f^{(5)}_{N,N}$'s:
\begin{eqnarray}
&& 120\, t^{1/2}\,  \,f^{(5)}_{1,1} \,\,
 =\,\,\, 64\cdot (K-\,E ) \, \, \nonumber \\
&&\quad -20\, \left(2\,t-3 \right) {K}^{3} \, -120\,E{K}^{2}\, +60\,{E}^{2}\, K
\nonumber \\
&&\quad + \left( 4\,t-5 \right)  \left( 2\,t-3 \right) {K}^{5}
 +15\, \left( 2\,t-3 \right)\,  {K}^{4}\, E\, +45\,{K}^{3}{E}^{2} \,
 \,  -15\,{K}^{2}{E}^{3} \nonumber \\
&& 360 \, t\, f^{(5)}_{2,2} \,\,\, =\, \,
149\, \left( t+2 \right)\,  K \,  -298\, \left( t+1 \right)\,  E\nonumber \\
&& \quad  +720\,{E}^{3}
 -90\, \left( {t}^{2}-2 \right) {K}^{3}\, \nonumber \\
&&\quad   +90\, \left( 2\,{t}^{2}-11\,t+2 \right) E{K}^{2} 
+1080\, \left( t-1 \right) {E}^{2}\, K \nonumber \\
&&\quad  +\left( 5\,{t}^{3}+28\,{t}^{2}-90\,t+60 \right) {K}^{5}\, 
 -10\, \Bigl( {t}^{3}-16\,{t}^{2}+24\,t  -4 \Bigr)\,  E{K}^{4}\,\nonumber \\
&& \quad  -5\, \left( 32\,{t}^{2}-179\,t+122 \right) {K}^{3}{E}^{2}\,
 \, -30\, \left( 19\,t-29 \right) {K}^{2}{E}^{3}\,
  -360\,{E}^{4}\, K\, \nonumber \\
&& 5400\, \,  t^{5/2} f^{(5)}_{3,3}\,  \, =\, \,\,  \, 
792\, \left( 4\,{t}^{2}+3\,t+8 \right)\,t  \cdot K
\, \,  -792\,t \left( 8\,{t}^{2}+7\,t+8 \right)\cdot  E \nonumber \\
&&\quad  -40\, \left( 72\,{t}^{4}-158\,{t}^{3}
+189\,{t}^{2}-156\,t+8 \right) {K}^{3} \nonumber \\
&&\quad  +240\, \left( 24\,{t}^{4}
-108\,{t}^{3}+29\,{t}^{2}-6\,t+4 \right)\, E{K}^{2} \nonumber \\
&&\quad  +120\, \left( 232\,{t}^{3}
-111\,{t}^{2}-180\,t-8 \right)\,  {E}^{2}\, K \nonumber \\
&&\quad  +160\, \left( t+1 \right)
 \left( 2\,{t}^{2}+103\,t+2 \right) {E}^{3}  \nonumber  \\
&&\quad +5\, \left( 96\,{t}^{5}-520\,{t}^{4}+1310\,{t}^{3}
-1589\,{t}^{2}+800\,t-88 \right)\,  {K}^{5} \nonumber \\
&&\quad  +5\, \left(424 -2488\,t+5051\,{t}^{2} -4962\,{t}^{3} 
+2008\,{t}^{4} -192\,{t}^{5}\right)\,  E\, {K}^{4} \nonumber \\
&&\quad  -5\, \left( 1984\,{t}^{4}-9228\,{t}^{3}+9423\,{t}^{2}
-3272\,t+816 \right)\,  {E}^{2}\, {K}^{3} \nonumber \\
&& \quad +5\, \left(784-4104\,t+11697\,{t}^{2}
-6056\,{t}^{3} \right)\,  {E}^{3}\, {K}^{2} \nonumber \\
&& \quad -40\, \left( 2\,{t}^{3}
+738\,{t}^{2}-567\,t+47 \right) {E}^{4}K \nonumber \\
&&\quad  +360\, \left( {t}^{2}-28\,t+1 \right) {E}^{5} \nonumber 
\end{eqnarray}

\subsection{ $f^{(6)}_{N,N}$}
Some $f^{(6)}_{N,N}$ read:
\begin{eqnarray}
&& 720\,  f^{(6)}_{1,1}\, \, = \, \, 
225\, \, -259\, \left(t -2 \right) {K}^{2} \,  -777\,K\, E 
 \nonumber \\
&&\quad-105\,{K}^{3}{E}^{3}+525\,{K}^{2}{E}^{2}
 +350\, \left(t-2 \right) \, E\, {K}^{3}\,
 +35\, \left( 6+{t}^{2}-6\,t \right) {K}^{4}\, \nonumber \\
&&\quad  -21\, \left( 6+{t}^{2}-6\,t \right)\,  E\, {K}^{5}\,\, 
 -105\, \left(t-2 \right)\, {E}^{2}{K}^{4}\,  \nonumber \\
&&\quad
- \left(t-2 \right)  \left( {t}^{2}-10\,t+10 \right)\,  {K}^{6}   \nonumber \\
&& 2160\, t\,  f^{(6)}_{2,2}   \, =\, \,  \,  2160\,t \, \,
  -544\, \left(15\,t-4 \right)\,  K\, E \, \nonumber \\
&&\quad   -1088\, \left( 1+t \right)\cdot  E^{2} \,
 -544\, \left( 6\,{t}^{2}-11\,t+2 \right)\cdot  K^{2}   \nonumber \\
&&\quad \, +50\, \left( 24\,{t}^{3}-98\,{t}^{2}+113\,t-36 \right) {K}^{4}\, 
-1200\,{E}^{4}\, +600\, \left( 9+t \right)\,  {E}^{3}\, K  \nonumber \\
&&\quad  +150\, \left(71\,t-60 \right) {K}^{2}{E}^{2} \, \, 
+100\, \left( 66+74\,{t}^{2}-157\,t \right) E\, {K}^{3}   \nonumber \\
&&\quad \, +360\,K\, {E}^{5}\,\,  \, 
-15\, \left(235\,t-264 \right) \, {K}^{3}\, {E}^{3} \nonumber \\
&&\quad  +3\, \left( 720\, -1889\,t\, 
+1490\,{t}^{2}\, -344\,{t}^{3}\right)\cdot  E\, {K}^{5}   \nonumber \\
&& \quad -90\, \left( 21+t \right) {E}^{4}{K}^{2}\,  \,  
-45\, \left( 92+74\,{t}^{2}-173\,t \right)\cdot  {E}^{2}{K}^{4} \,\nonumber \\
&&\quad   -3\, \left( 32\,{t}^{4}
-220\,{t}^{3}+504\,{t}^{2}-467\,t+150 \right)\,  {K}^{6} \nonumber
\end{eqnarray}

All these  $\,f^{(j)}_{N,N}$ displayed when expanded have their
leading coefficients starting as given in (\ref{2n}) and (\ref{2n-1}).
Let us give some $f^{(6)}_{N,N}$ as
a series to show the magnitude of the numerical coefficients involved.
The series expansion of  $f^{(6)}_{N,N}$ for the first values of $N$ reads:
\begin{eqnarray}
N=0:&&\quad \frac{t^{9}}{1073741824}
+\frac{37\,t^{10}}{8589934592}+\cdots \nonumber \\
N=1:&&\quad
\frac{7\,t^{12}}{4398046511104}
+\frac{21\,t^{13}}{2199023255552}\, +\cdots \nonumber \\
N=2:&&\quad \frac{21\,t^{15}}{1125899906842624}
+\frac{19215\,t^{16}}{144115188075855872}+\cdots \nonumber  \\
N=3:&&\quad \frac{10395\,t^{18}}{18446744073709551616}
+\frac{84315\,t^{19}}{18446744073709551616} +\cdots \nonumber  \\
N=4:&&\quad \frac{2335905\,t^{21}}{75557863725914323419136}
+\frac{166783617\,t^{22}}{604462909807314587353088}+\cdots  \nonumber 
\end{eqnarray}

\subsection{ $f^{(7)}_{1,1}$, $\, f^{(8)}_{1,1}$ and $\, f^{(9)}_{1,1}$}
Here we give the $f^{(j)}_{N,N}$ for the other values of $j=7,8,9$ and $N=1$:
\begin{eqnarray}
&& 5040\, t^{1/2}   \,   f^{(7)}_{1,1} \, \,  =\, \,   \, 
2304\cdot (K-E)\,  \,  \nonumber \\
&&\quad -784\, \left( 2\,t-3 \right)\cdot  {K}^{3} \, -4704\,{K}^{2}\, E\,
  +2352\,K\, {E}^{2}  \nonumber \\
&&\quad +840\, \left( 2\,t-3 \right)\,  {K}^{4}\, E 
  -840\,{K}^{2}\, {E}^{3}  \,\nonumber \\
&&\quad
  +56\, \left( 4\,t-5 \right)  \left( 2\,t-3 \right) {K}^{5}\, 
+2520\,{K}^{3}\, {E}^{2} \nonumber \\
&&\quad -28\, \left( 4\,t-5 \right)  \left( 2\,t-3 \right)\, E\, {K}^{6}\, 
  -210\, \left( 2\,t-3 \right)\,  {E}^{2}\, {K}^{5} \,\nonumber \\
&&\quad - \left(32\,{t}^{3}-156\,{t}^{2}+228\,t-105 \right) {K}^{7} \, 
  +105\,{K}^{3}{E}^{4}-420\,{K}^{4}{E}^{3}     \nonumber
\end{eqnarray}
\begin{eqnarray}
&& 645120 \, f^{(8)}_{1,1} \,  =\, \, \,
11025\,\,\, -38748\,K\,E \,\, \,
 -12916\, \left(t-2 \right)\, {K}^{2} \nonumber \\
&&\quad 
 +29610\,{K}^{2}{E}^{2} +1974\, \left( {t}^{2}-6\,t+6 \right)\, {K}^{4}\, 
 +19740\, \left( t-2 \right)\,  {K}^{3}\, E   \nonumber \\
&&\quad -8820\,{K}^{3}\, {E}^{3}\, \, 
-84\, \left( t-2 \right)  \left( {t}^{2}
-10\,t+10 \right)\,  {K}^{6}\,  \nonumber \\
&&\quad -1764\, \left( {t}^{2}-6\,t+6 \right)\,  {K}^{5}\, E \, 
-8820\, \left( t-2 \right)\,  {K}^{4}\, {E}^{2}\, \nonumber \\
&&\quad  +945\,{K}^{4}{E}^{4}\, \,  
 + \left( {t}^{4}-20\,{t}^{3}+48\,{t}^{2}
-56\,t+28 \right) {K}^{8}  \nonumber \\
&&\quad  +36\, \left( t-2 \right)  \left( {t}^{2}
-10\,t+10 \right)\,  {K}^{7}\, E\,\,\, 
+378\, \left( {t}^{2}-6\,t+6 \right) {K}^{6}{E}^{2} \nonumber \\
&&\quad  +1260\, \left( t-2 \right)\,  {K}^{5}\, {E}^{3}\,    \nonumber
\end{eqnarray}
\begin{eqnarray}
&& 362880 \, t^{1/2} \,f^{(9)}_{1,1} \,  =\, \,\,\,
 147456\cdot (K-E) \,\nonumber \\
&&\quad +157440\,K\, {E}^{2}\,
 -52480\, \left( 2\,t-3 \right)\cdot  {K}^{3}\, \,
-314880\,{K}^{2}\, E\,  \nonumber \\
&& \quad  \, -65520\,{K}^{2}\, {E}^{3} \, 
 +65520\, \left( 2\,t-3 \right) \,{K}^{4}\, E    \nonumber \\
&&\quad  +4368\, \left( 4\,t-5 \right)  \left( 2\,t-3 \right)\,  {K}^{5}
 \,  +196560\,{K}^{3}{E}^{2}   \nonumber \\
&& \quad  +12600\,{K}^{3}\, {E}^{4} -50400\,{K}^{4}{E}^{3}\, 
 -3360\, \left( 4\,t-5 \right)  
\left( 2\,t-3 \right)\, {K}^{6}\, E\, \nonumber \\
&&\quad +120\, \left( 105 -228\,t +156\,{t}^{2} 
-32\,{t}^{3} \right) \, {K}^{7} \,   \nonumber \\
&& \quad  -25200\, \left( 2\,t-3 \right) {K}^{5}{E}^{2} \, \, 
+630\, \left( 4\,t-5 \right) 
 \left( 2\,t-3 \right)\cdot  {K}^{7}\, {E}^{2}\,  \nonumber \\
&& \quad -945\,{E}^{5}\, {K}^{4}\,  +4725\,{E}^{4}\, {K}^{5}\, 
  +3150\, \left( 2\,t-3 \right)\, {K}^{6}\, {E}^{3} \nonumber \\
&& \quad +45\, \left( 32\,{t}^{3}
-156\,{t}^{2}+228\,t-105 \right) \,{K}^{8}\, E  \nonumber \\
&& \quad + \left( 128\,{t}^{4}
-960\,{t}^{3}+2460\,{t}^{2}-2572\,t+945 \right)\,  {K}^{9} \nonumber 
\end{eqnarray}

\section{Miscellaneous off diagonal $\, j$-particle contributions}
\label{D}

We display here some off diagonal $\, j$-particle contributions.
\begin{eqnarray}
\label{nice2}
&&2 \, s^2 \cdot C^{(2)}(0,2) \,  = \, \, 
2\,{s}^{2}\, \,  \, 
 -2\,\left( 1+{s}^{2} \right)\,{s}^{2} \cdot   K \, \nonumber \\
&&\quad+ \left(2\,{s}^{4}+{s}^{2} -2\right)\cdot  {K}^{2}\,  \, 
- \left( s-2 \right) 
 \left( s+2 \right) \,  K \, E \, \,   -2\,{E}^{2}   \nonumber \\
&&8\,  s^4 \cdot C^{(2)}(0,3) \, = \, \, 
{s}^{2} \left( 8+27\,{s}^{2}+8\,{s}^{4} \right) \, \,\nonumber \\
&&\quad -24\,{s}^{2} \left( 1+{s}^{2} \right)\cdot  E\, \, 
 -2\,{s}^{2} \left( 1+{s}^{2} \right) 
 \left( -4+13\,{s}^{2}+8\,{s}^{4} \right)\cdot  K   \nonumber \\
&&\quad+ \left( 1+{s}^{2} \right)  
\left( 8\,{s}^{8}+7\,{s}^{6}+3\,{s}^{4}\,
-8\,{s}^{2}-8 \right)\cdot   {K}^{2} \nonumber \\
&&\quad  +4\, \left( 4+6\,{s}^{2}+7\,{s}^{4}
+6\,{s}^{6} \right)\cdot  E\, K \,\,\,
  -8\, \left( {s}^{4}+1 \right)\cdot  {E}^{2}   \nonumber \\
&&18 \, s^6 \cdot C^{(2)}(0,4) \,  = \, 
 36\,\, ( 2+8\,{s}^{2}+13\,{s}^{4}+8\,{s}^{6}
+2\,{s}^{8})\,{s}^{2}  \nonumber \\
&&\quad  -24\,{s}^{2} \left( 1+{s}^{2} \right)  
\left( 6\,{s}^{8}+18\,{s}^{6}+11\,{s}^{4}
-12\,{s}^{2}-8 \right)\,  K \nonumber \\
&& \quad -48\,{s}^{2} \left( 1+{s}^{2} \right) 
 \left( 7\,{s}^{4}+15\,{s}^{2}+7 \right)\,  E \nonumber \\
&& \quad +\left( {s}^{2}+2 \right)  
\left( 72\,{s}^{12}+144\,{s}^{10}-60\,{s}^{8}
-200\,{s}^{6}+{s}^{4}+62\,{s}^{2}-16 \right)\,  {K}^{2}
 \nonumber \\
&& \quad + \left( 64-408\,{s}^{2}-576\,{s}^{4}+591\,{s}^{6}
+1088\,{s}^{8}+336\,{s}^{10} \right)\,  E\, K \nonumber \\
&&\quad  -4\, \left( 8-93\,{s}^{2}-200\,{s}^{4}
-93\,{s}^{6}+8\,{s}^{8} \right) {E}^{2} \nonumber \\
&&8 \, s^2 \cdot  C^{(2)}(1,2) \, \, = \, \,\, 7\,{s}^{2} \, \,
-4\, \left( 1+{s}^{2} \right)\cdot  E\, \nonumber \\
&&\quad  -2\, \left( 1+{s}^{2} \right)  \left( {s}^{2}-2 \right)\cdot  K\, 
\,,  \,     \nonumber \\
&&\quad  + \left( 1+{s}^{2} \right) 
 \left( 4 +3\,{s}^{2}  -5\,{s}^{4} \right)\,  {K}^{2}\,  \nonumber \\
&&\quad    +4\, \left({s}^{4}
-3\,{s}^{2}-3 \right) \, EK\, +8\,{E}^{2} \nonumber \\
&&6 \, s^4 \cdot C^{(2)}(1,3) \,  = \, \,  \, 9\,{s}^{4}\,\,  \, 
-4\, \left( 1+{s}^{2} \right) 
 \left( {s}^{4}+3\,{s}^{2}+1 \right)\cdot E  \nonumber \\
&& \,\quad 
 +4\, \left( 1+{s}^{2} \right)
  \left( 1\, +3\,{s}^{2}-{s}^{4} \right)\cdot  K\, \nonumber \\
&& \quad + \left( 10\, +8\,{s}^{2}-2\,{s}^{4}
-8\,{s}^{6}-5\,{s}^{8} \right) \cdot  {K}^{2} \nonumber \\
&& \quad + \left( -24-32\,{s}^{2}-13\,{s}^{4}
+16\,{s}^{6}+4\,{s}^{8} \right) \cdot E\, K \nonumber \\
&&\quad   +2\, \left( 7\,{s}^{4}+12\,{s}^{2}+7 \right)\cdot  {E}^{2} \nonumber
\end{eqnarray}

\section{Differential operators in the scaling limit}
\label{E}

The differential operators $\, L_{6}^{scal},\cdots ,
\, L_{10}^{scal}\,\, $ introduced in sec. \ref{scal}
read:
\begin{eqnarray}
&&L_6^{scal}  \, = \,\, \,64\,{x}^{6}\, Dx^{6}\, +320\,{x}^{5} \,  Dx^{5}\, 
-16\,{x}^{4} \left( 48+35\,{x}^{2} \right)\,  Dx^{4}\nonumber \\
&&\quad +32\,{x}^{3} \left( 91\,{x}^{2}-4 \right) \, Dx^{3}
+4\,{x}^{2} \left(848-1788\,{x}^{2}+259\,{x}^{4} \right)
 {{\it Dx}}^{2}\nonumber \\
&&\quad -20\,x \left( 80-380\,{x}^{2}+383\,{x}^{4} \right) 
\,  Dx\,\nonumber \\
&&\quad  -225\,{x}^{6}-2480\,{x}^{2}
+1600+17580\,{x}^{4} \nonumber 
\end{eqnarray}
\begin{eqnarray}
&&L_7^{scal}  \, = \, \,4\,{x}^{7} \, Dx^{7}
-56\, \left( 3+{x}^{2} \right)\,{x}^{5}  \, Dx^{5}
+8\,\left( 41+84\,{x}^{2} \right)\,{x}^{4} \, Dx^{4}\nonumber \\
&&\quad  +4\, \left(69\, -810\,{x}^{2}
+49\,{x}^{4} \right)\,{x}^{3}\, Dx^{3}\nonumber \\
&&\quad 
-8\,\left( 251-971\,{x}^{2}+372\,{x}^{4} \right)\,{x}^{2} \, Dx^{2}\nonumber \\
&&\quad 
+4\,\left( 275-2116\,{x}^{2}
+4212\,{x}^{4}\, -36\,{x}^{6} \right)\,x \,  Dx \nonumber \\
&&\quad 
-1100\, +2832\,{x}^{2}-35280\,{x}^{4}+1152\,{x}^{6}, \nonumber 
\end{eqnarray}
\begin{eqnarray}
&&L_8^{scal}  \, = \, \,  256\,{x}^{8} \, Dx^{8} \,
 -1024\,{x}^{7} \,  Dx^7\, 
-5376\, \left( 2 +{x}^{2} \right)\,{x}^{6}  \,  Dx^{6}\nonumber \\
&&\quad +256\, \left( 334
 +399\,{x}^{2} \right)\,{x}^{5}  \, Dx^5 \, \nonumber \\
&&\quad 
-32\,  \left(4040 +26304\,{x}^{2}
 -987\,{x}^{4} \right)\,{x}^{4}  \, Dx^4\, \nonumber \\
&&\quad -64\, \left(4216 -57384\,{x}^{2}
 +12027\,{x}^{4} \right)\,{x}^{3}  \, Dx^{3}\,\nonumber \\
&& \quad +16\, \left( 76688 -537424\,{x}^{2}+482478\,{x}^{4}\,
-3229\,{x}^{6} \right)\,{x}^{2} \, Dx^{2}\, \nonumber \\
&&\quad  -16\,\left(48400-598032\,{x}^{2} +2328262\,{x}^{4}
 -60013\,{x}^{6} \right)\, x \, Dx \, \nonumber \\
&&\quad +774400 -3342592\,{x}^{2}+72498272\,{x}^{4} 
-4879248\,{x}^{6}+11025\,{x}^{8}, 
  \nonumber 
\end{eqnarray}
\begin{eqnarray}
&&L_9^{scal}  \, = \, \,4\,{x}^{9}\,  Dx^{9}
+480\,{x}^{8}\, Dx^8\, \,  
-24\, \left( 5\,{x}^{2}-961 \right)\,{x}^{7} \, Dx^{7}
\nonumber \\
&&\quad +8\, \left( 71315-1092\,{x}^{2} \right)\,{x}^{6} \,  Dx^{6}\nonumber \\
&&\quad 
+12\,\left(645013 -19478\,{x}^{2}+91\,{x}^{4} \right)\,{x}^{5} \, Dx^{5}
\nonumber \\
&&\quad +8\, \left( 6985303+4920\,{x}^{4}
-354291\,{x}^{2} \right)\,{x}^{4} \,  Dx^{4} \nonumber \\
&&\quad +4\, \left(44460417 \, -3774790\,{x}^{2}\, 
+108828\,{x}^{4}-820\,{x}^{6} \right)\, {x}^{3} \, Dx^{3}
\nonumber \\
&&\quad -4\, \left(443021 +5872124\,{x}^{2}-382676\,{x}^{4} 
+9216\,{x}^{6} \right) \,{x}^{2}\,  Dx^{2} \nonumber \\
&&\quad 
+4\, \left( 576\,{x}^{8}-16128\,{x}^{6}+94812\,{x}^{4}
+9265148\,{x}^{2}-268975475 \right) \,x \,  Dx
\nonumber \\
&&\quad +1024\, \left(36\,{x}^{6}\,
 -2019\,{x}^{4} +66804\,{x}^{2}-1254400 \right) 
\nonumber
\end{eqnarray}
\begin{eqnarray}
&&L_{10}^{scal}  \, = \, \,1024\, {x}^{10}\, Dx^{10}
+168960\, {x}^{9}\, Dx^{9}\,
 +8448\,\left( 1368-5\,{x}^{2} \right)\, {x}^{8}  \, Dx^{8}
\nonumber \\
&&\quad +11264\, \left( 37880\, 
-399\,{x}^{2} \right)\, {x}^{7} \, Dx^{7}\nonumber \\
&&\quad 
+4224\, \Bigl(2194904 \, 
 -44164\,{x}^{2}+133\,{x}^{4}\Bigr)\,{x}^{6} \,   Dx^{6}\nonumber \\
&&\quad 
 +128\, \left( 946138408-30108276\,{x}^{2}+259215\,{x}^{4}
 \right)\,{x}^{5} \, Dx^{5} \nonumber \\
&&\quad +32\, \left(29025917984\, -1305848840\,{x}^{2}
 +21377796\,{x}^{4} -86405\,{x}^{6}  \right) \, {x}^{4} \, Dx^{4}
\nonumber \\
&&\quad +64\, \left( 60403578784-3569603544\,{x}^{2} \,
 +92712956\,{x}^{4}-1057221\,{x}^{6} \right)\,{x}^{3} \,  Dx^{3}
\nonumber \\
&&\quad +4\,\Bigl(1794785734400 -134201812672\,{x}^{2} +5056843872\,{x}^{4}
 \nonumber \\
&& \quad -110074968\,{x}^{6}\, +1057221\,{x}^{8} \Bigr)\, {x}^{2} \,  Dx^{2}\nonumber \\
&&\quad  +972\, \Bigl(3056659200 -330174912\,{x}^{2} 
+ 18778592\,{x}^{4}\, -694968\,{x}^{6} \nonumber \\
&&\quad  +18375\,{x}^{8} \Bigr)\cdot  x \, Dx\, \quad 
\, \, \,  -893025\,{x}^{10}  \nonumber 
\end{eqnarray}

\end{document}